\documentclass[12pt]{cernprep}
\usepackage{graphicx}
\usepackage{amsmath}
\usepackage{amssymb}
\usepackage{epsfig,color,rotating}
\usepackage{epstopdf}
\usepackage{thumbpdf}
\usepackage[pdftex,
        pdftitle={A Document With An Image},
        pdfauthor={Udo K. Schuermann, Ringlord Technologies},
        pdfsubject={A Demonstration with Extra Cheese},
        pdfkeywords={latex pdf gimp batch thumbnails hyperlinks},
        pagebackref,
        pdfpagemode=None,
        bookmarksopen=true]{hyperref}
\usepackage{euscript}
\usepackage{multirow}
\usepackage{colortbl}
\usepackage{calc}

\makeatletter
\newlength\twolinebox@linelength
\newlength\twolinebox@columnheight
\newcommand{\twolinebox}[2]{%
   \setlength{\twolinebox@linelength}%
             {\maxof{\widthof{#1}}{\widthof{#2}}}%
   \setlength{\twolinebox@columnheight}{\heightof{#1}+\depthof{#1}+0.2em+0.4em/2+\heightof{0}/2}%
   \raisebox{0pt}[\twolinebox@columnheight][\heightof{\vbox{\vskip0.2em\hbox to 
   \twolinebox@linelength {#1\hfil}\vskip0.4em\hbox to 
   \twolinebox@linelength {#2\hfil}}}+\depthof{\vbox{\vskip0.2em\hbox to 
   \twolinebox@linelength {#1\hfil}\vskip0.4em\hbox to 
   \twolinebox@linelength {#2\hfil}}}-\twolinebox@columnheight+0.2em]{\vbox to 
   \twolinebox@columnheight{\vskip0.2em\hbox to 
   \twolinebox@linelength {#1\hfil}\vskip0.4em\hbox to 
   \twolinebox@linelength {#2\hfil}}}%
}
\newcommand{\pT} {\mbox{$p_{\rm T}$}}
\newcommand{\pTl} {\mbox{$p_{\rm T,l}$}}
\newcommand{\pTlp} {\mbox{$p_{{\rm T},{\rm l}^+}$}}
\newcommand{\pTlm} {\mbox{$p_{{\rm T},{\rm l}^-}$}}
\newcommand{\pTll} {\mbox{$p_{\rm T,ll}$}}
\newcommand{\kT} {\mbox{$k_{\rm T}$}}
\newcommand{\GeVc} {\mbox{GeV/$c$}}
\newcommand{\MeVc} {\mbox{MeV/$c$}}

\newcommand{\MeVcsq} {\mbox{MeV/$c^2$}}
\newcommand{\MW} {\mbox{$M_{\rm W}$}}

\newcommand{\MWp} {\mbox{$M_{{\rm W}^+}$}}
\newcommand{\MWm} {\mbox{$M_{{\rm W}^-}$}}

\newcommand{\MZ} {\mbox{$M_{\rm Z}$}}   
\newcommand{\Wpm} {\mbox{W$^\pm$}} 
\newcommand{\Wp} {\mbox{W$^+$}}
\newcommand{\Wm} {\mbox{W$^-$}}  
\newcommand{\Zp} {\mbox{Z$^+$}}
\newcommand{\Zm} {\mbox{Z$^-$}}  
\newcommand{\etal} {\mbox{$\eta_{\rm l}$}}
\newcommand{\etalp} {\mbox{$\eta_{{\rm l}^+}$}}
\newcommand{\etalm} {\mbox{$\eta_{{\rm l}^-}$}}
\newcommand{\lp} {\mbox{l$^+$}}
\newcommand{\lm} {\mbox{l$^-$}}
\newcommand{\lpm} {\mbox{l$^\pm$}}

\newcommand{\Mll} {\mbox{$M_{\rm l l}$}}
\newcommand{\yll} {\mbox{$y_{\rm l l}$}}
\newcommand{\dbar} {\mbox{$\bar d$}}
\newcommand{\ubar} {\mbox{$\bar u$}}
\newcommand{\sbar} {\mbox{$\bar s$}}
\newcommand{\cbar} {\mbox{$\bar c$}}
\newcommand{\bbar} {\mbox{$\bar b$}}

\newcommand{\uv} {\mbox{$u_{\rm v}$}}
\newcommand{\dv} {\mbox{$d_{\rm v}$}}
\newcommand{\GamW} {\mbox{$\Gamma_{\rm W}$}}
\newcommand{\GamWp} {\mbox{$\Gamma_{{\rm W}^+}$}}
\newcommand{\GamWm} {\mbox{$\Gamma_{{\rm W}^-}$}}

\newcommand{\GamZ} {\mbox{$\Gamma_{\rm Z}$}}
\newcommand{\pp} {\mbox{pp}}
\newcommand{\ppbar} {\mbox{p\={p}}}
\newcommand{\dd} {\mbox{dd}}

\newcommand{\rhol} {\mbox{$\rho_{\rm l}$}}

\newcommand{\FlatAsymW} {\mbox{$\cal{A}_{\rm W}$}}
\newcommand{\FlatAsymZ} {\mbox{$\cal{A}_{\rm Z}$}}   
                           
\newcommand{\RWZ} {\mbox{$\cal{R}_{\rm WZ}$}}      
\newcommand{\RWZmod} {\mbox{$\cal{R}_{\rm WZ}^{\rm mod}$}}
\newcommand{\RWZQCD} {\mbox{$\cal{R}_{\rm WZ}^{\rm QCD}$}}
\newcommand{\RZnorm} {\mbox{$\cal{R}_{\rm Z}^{\rm norm}$}} 
\newcommand{\Asym} {\mbox{$\cal{A}$}}
\newcommand{\DAsym} {\mbox{$\cal{D}$}}
\newcommand{\ratC} {\mbox{$\cal{C}$}}

\newcommand{\tw}      {\textwidth}

\newcommand\WINHAC[0] {\textsf{WINHAC}}
\newcommand\ZINHAC[0] {\textsf{ZINHAC}}

\newcommand\Pythia[0] {\textsc{Pythia}}

\def\slashii#1{\setbox0=\hbox{$#1$}            
  \dimen0=\wd0                                 
  \setbox1=\hbox{\sl/} \dimen1=\wd1            
  \ifdim\dimen0>\dimen1                        
     \rlap{\hbox to \dimen0{\hfil\sl/\hfil}}   
     #1                                        
  \else                                        
     \rlap{\hbox to \dimen1{\hfil$#1$\hfil}}   
     \hbox{\sl/}                               
  \fi}

\definecolor{rltbrightred}{rgb}{1,0,0}
\definecolor{rltred}{rgb}{0.75,0,0}
\definecolor{rltdarkred}{rgb}{0.5,0,0}
\definecolor{rltbrightgreen}{rgb}{0,0.75,0}
\definecolor{rltgreen}{rgb}{0,0.5,0}
\definecolor{rltdarkgreen}{rgb}{0,0,0.25}
\definecolor{rltbrightblue}{rgb}{0,0,1}
\definecolor{rltblue}{rgb}{0,0,0.75}
\definecolor{rltdarkblue}{rgb}{0,0,0.5}
\definecolor{webred}{rgb}{0.5,.25,0}
\definecolor{webblue}{rgb}{0,0,0.75}
\definecolor{webgreen}{rgb}{0,0.5,0}
\definecolor{Black}{rgb}{0,0,0}
\definecolor{Greymax}{rgb}{0.65,0.65,0.65}
\definecolor{Greycen}{rgb}{0.75,0.75,0.75}
\definecolor{Greymin}{rgb}{0.85,0.85,0.85}
\definecolor{hl}{rgb}{
                0.909803922,       
                0.82745098,               
                0.909803922}
\begin{document}
%
\begin{titlepage}
\docnum{CERN--PH--EP/2010-007 }
\docnum{TPJU--1/2010}
\date{12 March 2010} 
\vspace{1cm}
\begin{center}
\title{\Large{$\mathbf{\Delta M_{\rm W} \leq 10}$~MeV/$\mathbf{c^2}$ at the LHC: 
a forlorn hope?$\,^\dagger$}}
\end{center}

\vspace*{2cm}

\begin{abstract} 
At the LHC, the measurement of the W mass with a precision of  $\cal{O}$(10)~\MeVcsq\  is both mandatory and difficult. In the analysis strategies proposed so far, shortcuts have been made that are justified for proton--antiproton collisions at the Tevatron, but not for proton--proton collisions at the LHC. The root of the problem lies in the inadequate knowledge of parton density functions of the proton. It is argued that in order to reach a 10~\MeVcsq\ precision for the W mass, more precise parton density functions of the proton are needed, and an LHC-specific analysis strategy ought to be pursued. Proposals are made on both issues.
\end{abstract}

\vfill  \normalsize
\begin{center}
M.W.~Krasny$^{1,\star}$,
F.~Dydak$^2$,
F.~Fayette$^1$, 
W.~P\l{}aczek$^3$, and
A.~Si\'{o}dmok$^{1,3}$

\vspace*{5mm} 

$^1$~LPNHE, Universit\'{e}s Paris VI et VII and CNRS--IN2P3, Paris, France \\
$^2$~CERN, Geneva, Switzerland \\ 
$^3$~Institute of Physics, Jagiellonian University, Cracow, Poland \\

\vspace*{5mm}

\submitted{(To be submitted to European Physical Journal C)}
\end{center}

\vspace*{5mm}
\rule{0.9\textwidth}{0.2mm}

\begin{flushleft}
\begin{footnotesize}
\hspace{15mm}
$^\dagger$ Work in part supported by the cooperation programme 
between the French IN2P3 and Polish \\
\hspace{15mm} 
COPIN Laboratories No.\ 05-116, 
and by the EU Marie Curie Research Training Network grant \\ 
\hspace{15mm}
No.\ MRTN-CT-2006-035505 \\
\hspace{15mm} 
$^\star$~Corresponding author; e-mail: krasny@lpnhep.in2p3.fr
\end{footnotesize}
\end{flushleft}

\end{titlepage}

\newpage
\mbox{ }

%

\section{Introduction}
\label{INTRODUCTION}

In much the same way as precise measurements of radiative corrections served
to test and establish QED, precise measurements of input parameters and their use in
the calculation of radiative corrections in the Electroweak Standard Model
serve as benchmarks for new theoretical concepts.
Therefore, besides the direct searches for new phenomena, the precision measurement
of parameters of the Electroweak Standard Model\footnote{Hereafter referred to as `electroweak parameters'.} ---e.g., the W mass---with greater precision than available from LEP and the Tevatron, is an important and indispensable part of the LHC programme.

Whilst the Z mass (\MZ ) is well measured to $\pm 2.1$~\MeVcsq~\cite{Amsler:2008zzb},
\MW\ is measured at the 
Tevatron to $\pm 31$~\MeVcsq~\cite{mW-Tevatron}\footnote{The ultimate W mass error
at the Tevatron may be as low as $\pm 15$~\MeVcsq .}
and at LEP to $\pm 33$~\MeVcsq~\cite{mW-LEPII}. Of the three independent input 
parameters of the Electroweak Standard model,  \MW , \MZ\ and the fine-structure constant,
\MW\ is by one order of magnitude less precise than \MZ\ that is second-best. 

Although a precision
of \MW\ that matches the precision of \MZ\ is experimentally
not within reach, a much better precision than available today is 
desirable to exploit the full potential 
of the relation between \MW\ and the Fermi coupling 
constant $G_{\rm F}$ that is also 
well measured with a relative precision of $1 \times 10^{-5}$.

The relation between $G_{\rm F}$ and the three input parameters,  
\MW , \MZ\ and the fine-structure constant, is a cornerstone of the
Electroweak Standard Model. Radiative corrections of this relation that depend
{\em inter alia\/} on the mass of the Higgs boson, 
suggest a broad range for the Higgs mass that is nevertheless 
well within reach at the LHC.
However, in case the Higgs boson will not be found, the hunt for
alternative models of electroweak symmetry breaking will be on. 
Then the highest possible precision of \MW\ will be a central issue, for
a better measured relation between the
quantities $G_{\rm F}$, \MW , \MZ , and the fine-structure constant,
will put more stringent constraints on theoretical models. 
 
In previous analyses, it was claimed that an \MW\ precision
of 10~\MeVcsq\ or better will be obtained at the LHC~\cite{mW-ATLAS, mW-CMS}.
This paper questions such claims and argues that shortcuts have been
made that are not justified, and hence the claimed measurement precision 
is much too optimistic. The reason is that the analysis of \pTl\
spectra from leptonic W and Z boson decays in 
\ppbar\ collisions at the Tevatron---that served as template for the respective 
analyses at the LHC---benefits from symmetry properties
that are absent in \pp\ collisions at the LHC. A
considerably better knowledge of the
$u_{\rm v} - d_{\rm v}$, $s - c$, and $b$ parton density functions (PDFs) of the 
proton\footnote{Throughout this
paper, PDFs refer to the proton.} than available today is needed, together with 
an LHC-specific measurement and analysis programme. 

No improvement of the current situation
is expected unless special experimental efforts are made to obtain the
missing high-precision PDFs. Two ways forward are discussed.
One is to complement the \pp\ programme of the LHC with a deuteron-deuteron
collision programme. Another is to obtain missing input from 
a new high-precision muon--nucleon scattering experiment, and to analyze
these data coherently with LHC \pp\ and Tevatron \ppbar\ data.

This paper is structured as follows. 
Section~\ref{WMASSATTHELHC}
discusses the subtleties of the W mass measurement at the LHC, with emphasis
on biases caused by the inadequate knowledge of certain PDFs.

Section~\ref{LHCSPECIFICPROGRAMME} describes the salient features of an
LHC-specific programme for the precision measurement of electroweak
parameters at the LHC.

In Section~\ref{WAYSFORWARD} two experimental programmes
are put forward that would permit a 10~\MeVcsq\ precision of \MW\ at the LHC.

\section{Measurement of the W mass at the LHC}
\label{WMASSATTHELHC}

Throughout this paper, it is taken for granted that the intrinsic   
\Wp\ and \Wm\ masses are equal\footnote{The best experimental support  
of this assertion stems from a comparison of the measured $\mu^+$ and 
$\mu^-$ lifetimes~\cite{Amsler:2008zzb},
which translates into an equality of \Wp\ and
\Wm\ masses at the 1.6~\MeVcsq\ level, 
a precision which is out of reach at the LHC.}.

\subsection{The lepton transverse momentum and the scale gap}
 
In \pp\ as well as in \ppbar\ collisions, \MW\ is determined by the
Jacobian peak in the \pT\ spectrum of charged 
leptons from ${\rm W} \rightarrow {\rm l}\nu$
decays. The scale gap between the Jacobian peak around 40~GeV/{\it c}  
and the wanted \MW\ precision at the 10~\MeVcsq\ level 
amounts to a factor of 4000. 
 
\begin{figure}[h]
\begin{center}
\vspace*{3mm}
\includegraphics[width=0.8\textwidth]{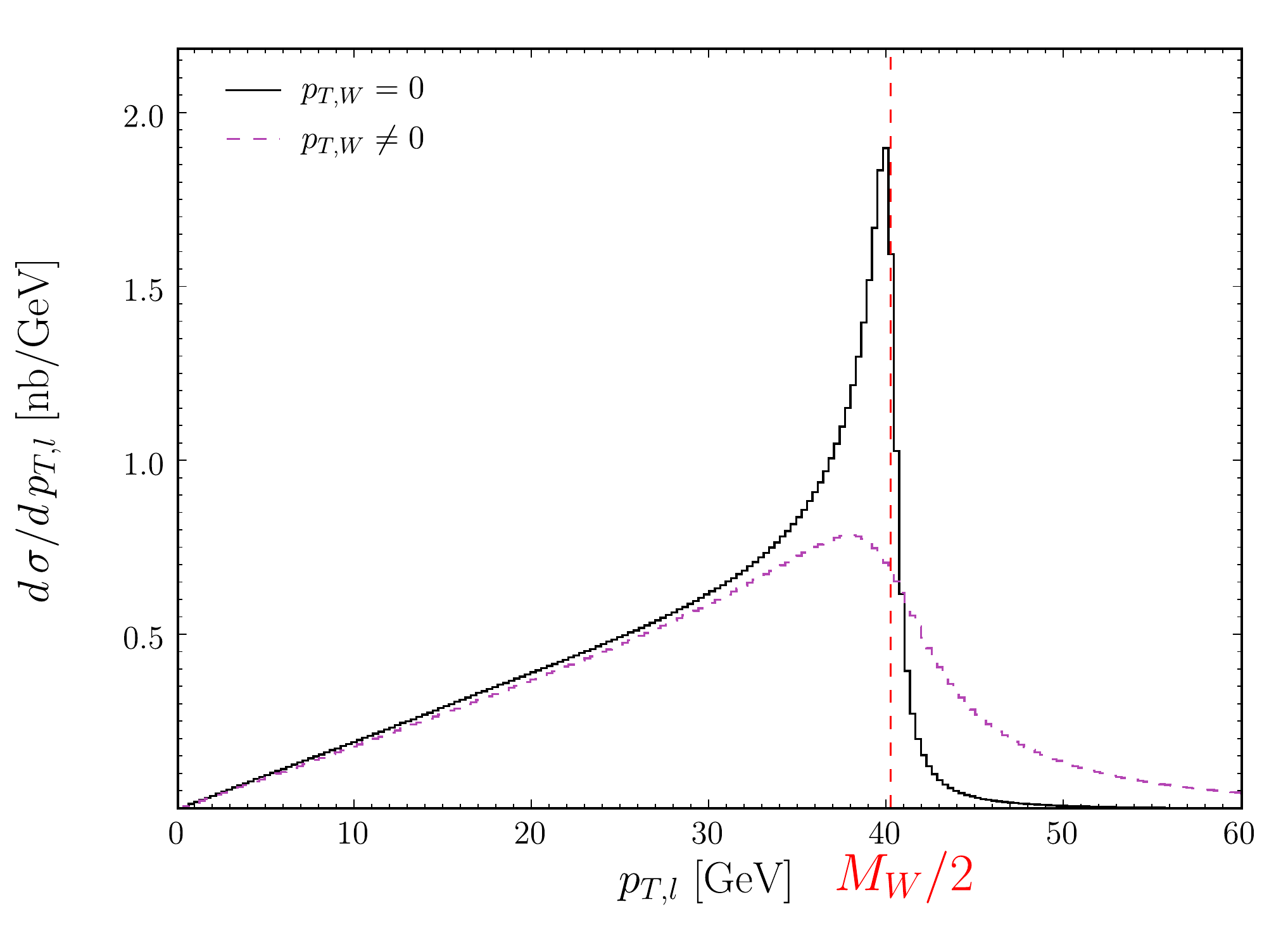}
\end{center}
\vspace*{-8mm}
\caption{Simulation of the $p_{\rm T}$ spectrum of charged leptons 
from the decay ${\rm W} \rightarrow {\rm l}\nu$; the full line is generated
with zero $p_{\rm T}$ of the W's, the broken line represents the
effect of a non-zero $p_{\rm T}$ as predicted by PYTHIA.}
\label{pTspectrumofW}
\end{figure}

The quantitative consequences of this scale gap are 
highlighted in Fig.~\ref{pTspectrumofW} which shows 
the change of the \pT\  spectrum of
charged leptons from the decay ${\rm W} \rightarrow {\rm l}\nu$ by the
inclusion of what PYTHIA~\cite{PYTHIA} predicts as \pT\ of W's at the
LHC. 
Since the W mass depends on the characteristics of the
Jacobian peak, it is intuitively clear that a very precise 
understanding of the shape of the 
$p_{\rm T}$ spectrum is mandatory, as is the calibration of 
the relation between the \pT\ spectrum of W decay leptons 
and \MW\ by means of the reference relation between the \pT\ spectrum 
of Z decay leptons and the precisely known \MZ .  
Great care must be 
devoted to all effects that cause either  
the production characteristics of W and Z to be different, 
or the decay characteristics of ${\rm W} \rightarrow {\rm l}\nu$ 
and ${\rm Z} \rightarrow {\rm l}^+{\rm l}^-$ to be different, or both. 
Any such difference would lead to different $p_{\rm T}$ spectra 
of leptons from W decays and from the reference Z decays.

The detector acceptance needs consideration, too. Since
charged leptons with a pseudorapidity $|\eta| > 2.5$ can hardly be 
measured, this limitation of the pseudorapidity range impacts on the 
\pT\ spectrum of charged leptons.

\subsection{Parton density functions}

Table~\ref{contributingquarks} recalls that quite different 
quark--antiquark pairs contribute to the 
production of \Wp , \Wm\ and Z. 
\begin{table}[h] 
\begin{center}
\vspace*{3mm}
\begin{tabular}{|c|l|} 
\hline                                     
\Wp\  & ${\rm u}\bar{\rm d} + {\rm u}\bar{\rm s} + {\rm u}\bar{\rm b} + {\rm c}\bar{\rm d} 
            + {\rm c}\bar{\rm s} + \cdots $ \\
\Wm\ & ${\rm d}\bar{\rm u} + {\rm d}\bar{\rm c} + {\rm s}\bar{\rm u} + {\rm s}\bar{\rm c} 
            + \cdots $ \\
Z        & ${\rm u}\bar{\rm u} + {\rm d}\bar{\rm d} + {\rm s}\bar{\rm s} + {\rm c}\bar{\rm c} 
            + {\rm b}\bar{\rm b}  + \cdots $ \\
\hline
\end{tabular}
\vspace*{2mm}
\caption{Quark--antiquark pairs that contribute to \Wp , \Wm\ and Z production.}
\label{contributingquarks}
\end{center}
\end{table}
The following properties of the contributing quarks lead to 
intrinsic differences in the $p_{\rm T}$ spectra of leptons from \Wp , \Wm\ and Z 
leptonic decays: 
(i) their PDFs, and 
(ii) their weak coupling constants.

In this paper, in contrast to the usual nomenclature where a PDF is a one-dimensional function
of the variable $x$, 
where $x$ denotes the fractional longitudinal momentum of the respective
parton of the proton longitudinal momentum, what is termed `PDF' generally refers to a two-dimensional function
of $x$ and \kT , where \kT\ is the transverse momentum of the respective parton, unless explicitly
specified otherwise.
The concept of a two-dimensional PDF is motivated by the large 
transverse momentum of the annihilating
quarks in the production of W and Z which is not at the few 100~\MeVc\ level like for
the production of low-mass particles such as pions, but rather at the level of several \GeVc .
The differential of the two-dimensional PDF of the quark q, ${\rm d}  q(x, k_{\rm T}; Q^2)$, 
denotes the number d$N$  of quarks of type q 
with a fraction of the proton longitudinal momentum in the range $[x, x+{\rm d}x]$,
with a transverse momentum in the range $[k_{\rm T},  k_{\rm T}+{\rm d} k_{\rm T}]$, at the scale $Q^2$. The one-dimensional PDF $q(x; Q^2)$, referred to below as `\kT -integrated PDF',
is the integral of the two-dimensional PDF $q(x, k_{\rm T}; Q^2)$ over \kT . 
Whereas the two-dimensional PDF has one longitudinal and one transverse dimension, 
the \kT -integrated PDF has only the longitudinal dimension.

The concept of two-dimensional PDFs takes into account (i)
the correlation between 
$x$ and \kT\ of the contributing quarks and antiquarks 
(small $x$ is correlated with large $k_{\rm T}$), (ii)
the correlation of \kT\  with the hardness scale of the 
process (the W and Z masses are different), and (iii)  
the dependence of \kT\ on the quark type (heavier quarks 
have larger \kT ).

The  two-dimensional PDFs 
are not process-independent universal functions\footnote{This is a departure from
the conventional approach that considers PDFs as universal, i.e. process-independent.}. Their functional forms are not predicted by QCD,
they are experimentally determined.
Their use is restricted to the analysis of purely leptonic observables
for which the initial- and final-state interactions can be factorized\footnote{There is no gluon exchange between initial and final state.}.
 
Throughout this paper, the two-dimensional PDFs refer to the 
scale $Q^2 = M_{\rm W}^2$.

Five quark flavours participate in the production of W and Z. Since quarks and antiquarks 
are to be considered, {\em a priori\/} ten two-dimensional PDFs need to be known.

\subsection{W and Z polarization}

By virtue of the different weak coupling constants 
and the different longitudinal and transverse momentum distributions 
of the annihilating quarks 
in W and Z production, 
the spin components in the longitudinal and transverse
directions and the polarizations, respectively, are different for \Wp , \Wm\ and Z. 
Because the W mass is determined from the \pT\ spectrum
of decay leptons, the interest focuses on the direction 
{\it perpendicular\/} to the beam.
The respective non-zero 
spin components perpendicular to the beam
direction constitute `longitudinal' 
polarizations\footnote{In analogy 
to the longitudinal polarization vector of a virtual 
photon.}.

The differences in the
longitudinal 
\Wp , \Wm\ and Z polarizations propagate   
through leptonic-decay characteristics into differences of 
the charged-lepton \pT\ spectra.

\subsection{\Wp , \Wm\  and Z leptonic-decay characteristics}

With respect to the \Wp , \Wm\ and Z spin directions, 
the angular distributions
of decay leptons are different according to the V$-$A and V$+$A
amplitudes in the boson--lepton coupling. In the 
W$^{\pm}$ rest frame, the pure V$-$A amplitude leads to the following 
angular distribution
of the charged-lepton emission amplitude:
\begin{equation}
w(\theta) \propto 1 \pm \cos{\theta^\ast}   \; ,
\end{equation}
where $\theta^\ast$ denotes the angle between the direction of the spin vector
and charged-lepton emission. In the Z rest frame, the 
mixture of V$-$A and V$+$A amplitudes\footnote{Because of Nature's choice 
of $\sin^2 \theta_{\rm w}$ close to 1/4, the V$-$A and V$+$A amplitudes
are nearly equal and the Z decay is nearly parity-conserving, in contrast to W decay
which violates parity maximally.}
leads to the angular distribution
\begin{equation}
w(\theta) \propto 1 + \gamma \cos{\theta^\ast}   \; ,
\end{equation}
where $0 < \gamma \ll 1$. 

The charged-lepton emission asymmetries are modified by the Lorentz boost
from the boson rest frame into the laboratory system.

On top of the genuine differences in the 
longitudinal polarizations 
of \Wp , \Wm\  and Z bosons, an important contribution to  
the differences in the \pT\ distribution of charged leptons
in the laboratory system
stems from the interference between transverse and
longitudinal boson polarization amplitudes.

Altogether, from the different longitudinal polarizations of \Wp , 
\Wm\ and Z, in conjunction with their different angular
distributions of charged-lepton emission, and 
in conjunction with their momentum spectra, 
the question arises whether the differences of the \pT\ spectra of decay leptons
from \Wp , \Wm\  and Z  can be sufficiently well understood to 
overcome the scale gap.

\subsection{Shortcuts revisited}

There are important differences of \Wp , \Wm\  and Z
production in \pp\ collisions at the LHC and in \ppbar\
collisions at the Tevatron. 
 
In \ppbar\ collisions at the Tevatron, there is a small 
forward--backward asymmetry in the production of charged 
leptons from Z decay, and a strong asymmetry from the decays of \Wp\
and \Wm , since e.g. \Wp\ are produced
preferentially along the incoming proton direction.
However, the rates and the momentum spectra of positive 
leptons from \Wp\ at the polar angle $\theta$ are 
the same as the rates of negative leptons from \Wm\ 
at the polar angle $\pi - \theta$. The same holds when
integrated over the same range of $\theta$ and $\pi - \theta$,
respectively. This lends itself to a common analysis of leptons 
with positive and negative charge.
 
In \pp\ collisions at the LHC, there is for any of the three bosons 
forward--backward symmetry in the production of charged leptons: 
at the polar angles $\theta$ and $\pi - \theta$, the 
rates and the momentum spectra are identical. However, 
the rates and the momentum spectra are mutually different between  
\Wp , \Wm\ and Z. In particular, the difference
in the rates and the momentum spectra of charged leptons from 
\Wp\ and \Wm\ decays renders  
a common analysis of leptons with positive and negative 
charge questionable.

\begin{figure}[h]
\begin{center}
\begin{tabular}{cc}
\includegraphics[height=0.35\textwidth]{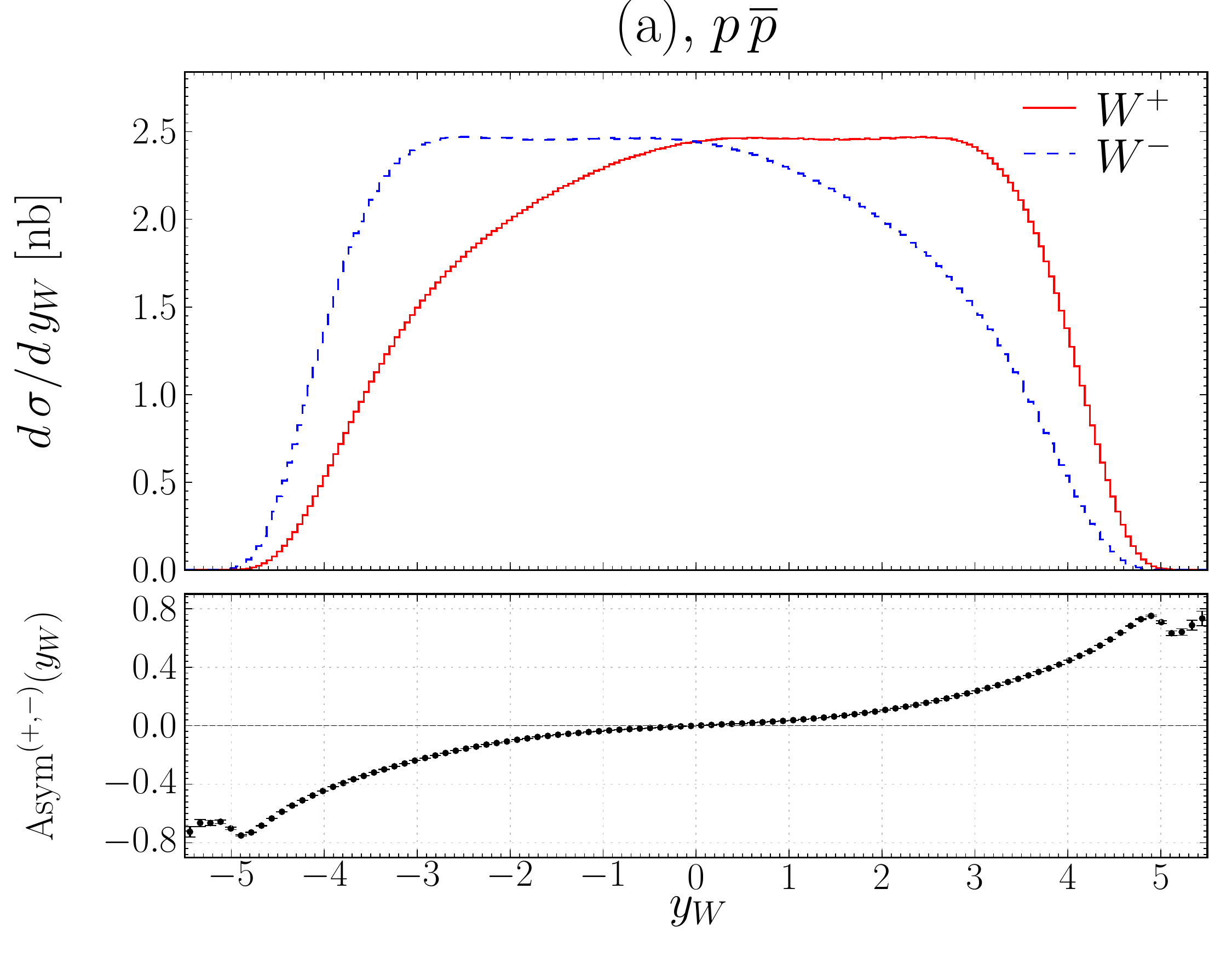} &
\includegraphics[height=0.35\textwidth]{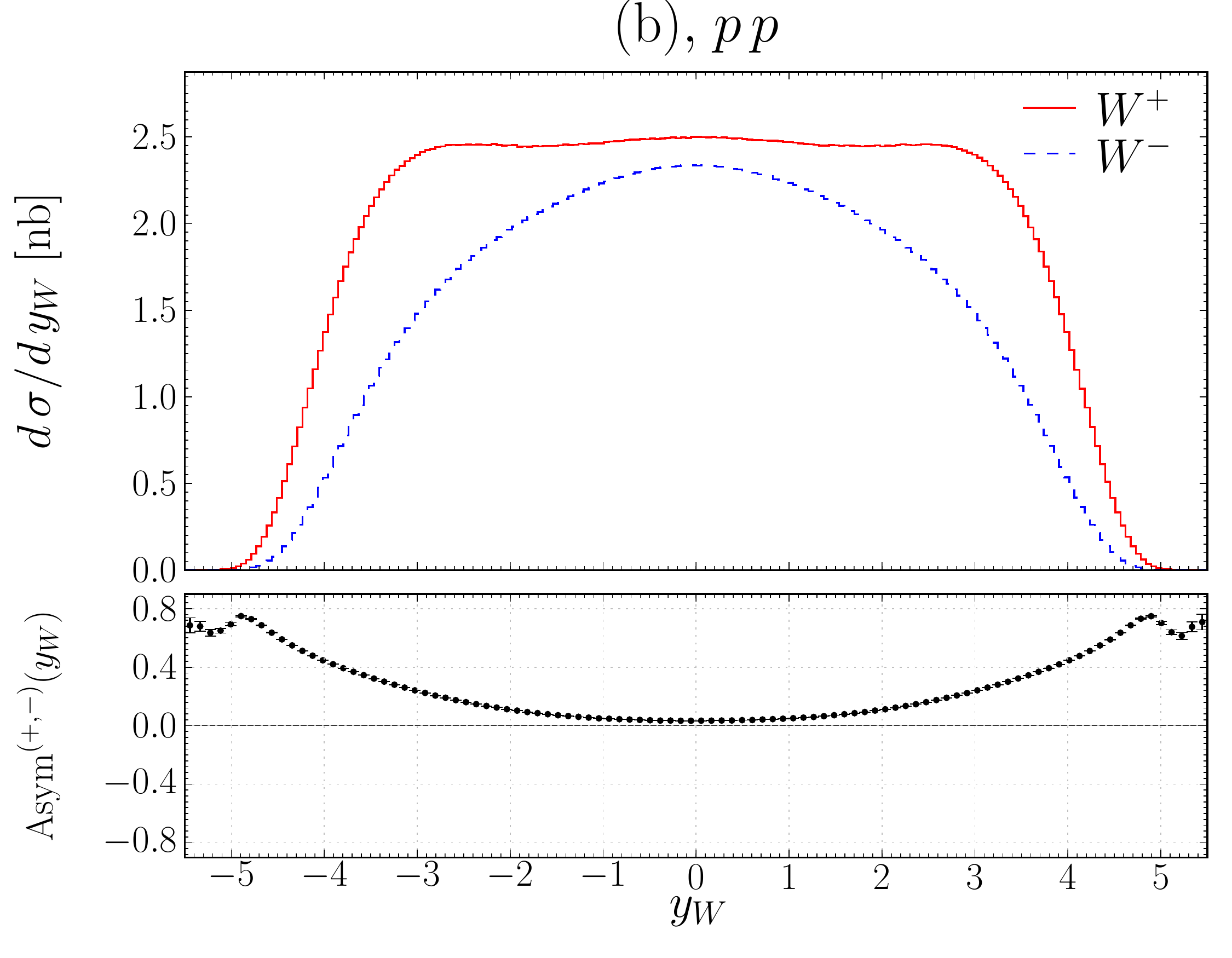} \\
\includegraphics[height=0.35\textwidth]{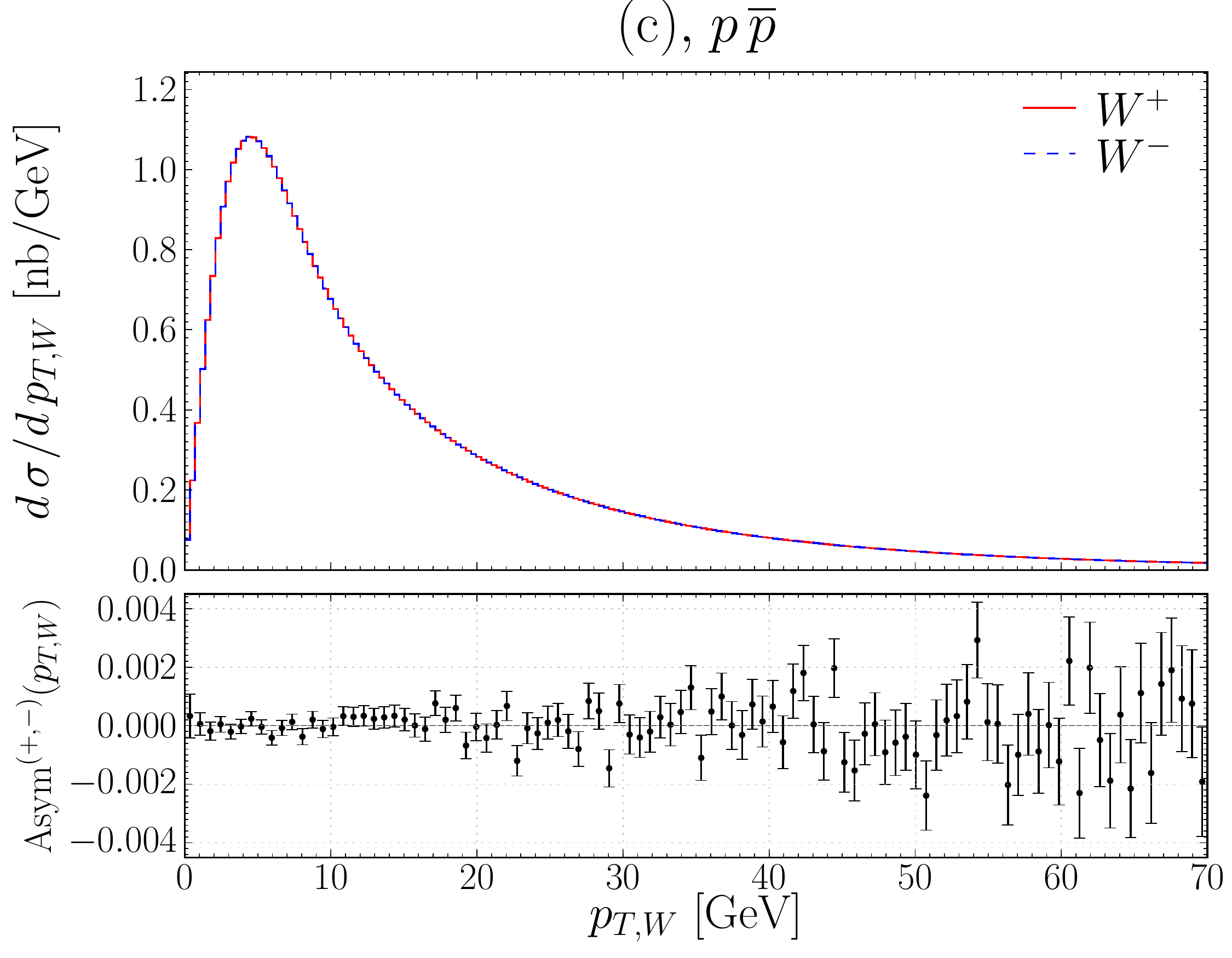} &
\includegraphics[height=0.35\textwidth]{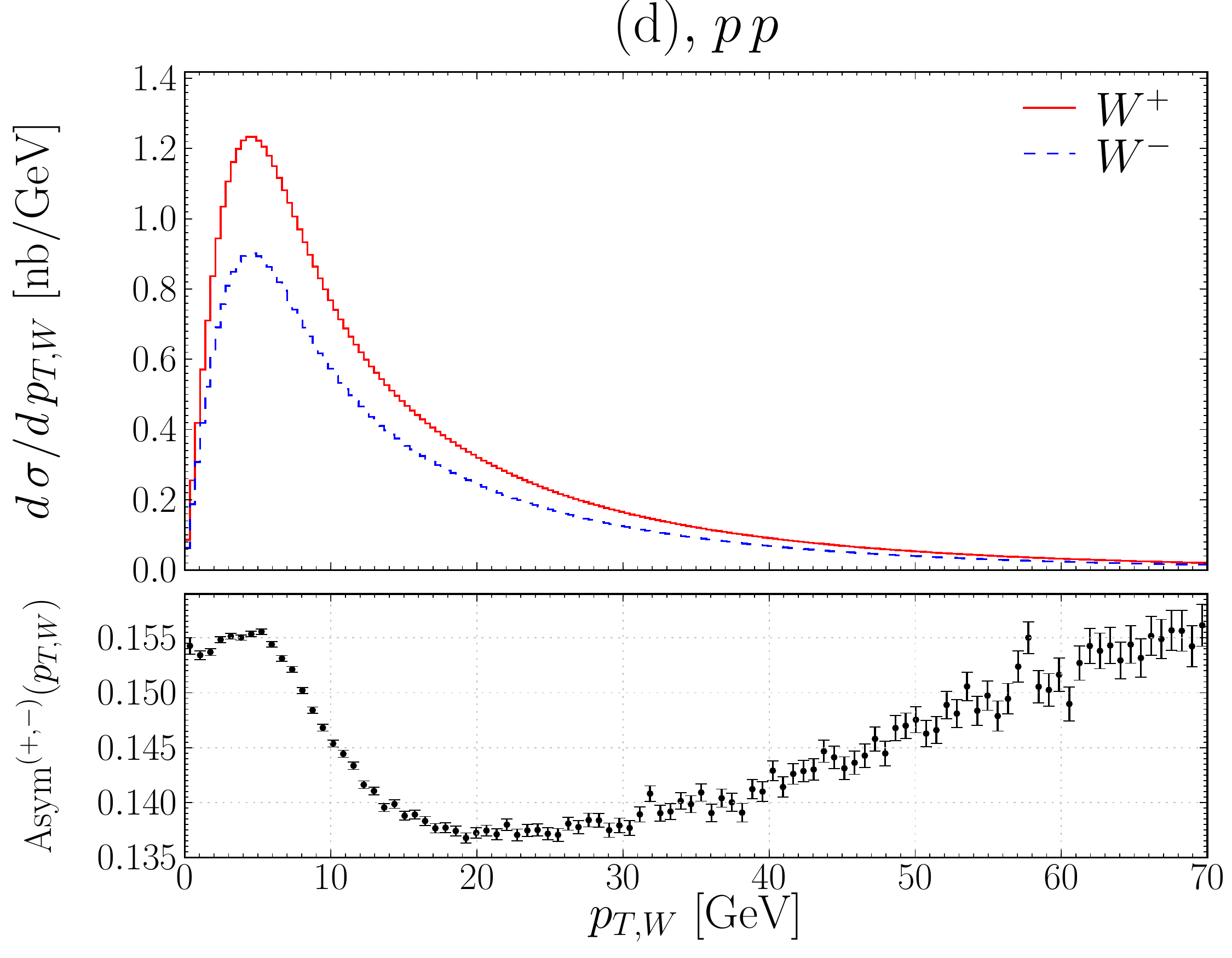} \\
\end{tabular}
\end{center}
\caption{Rapidity $y$ and the transverse momentum
\pT\ of \Wpm 's, in \ppbar\ collisions (left panels)
and in \pp\ collisions (right panels).}
\label{WyandpT}

\end{figure}

\begin{figure}[h]
\begin{center}
\begin{tabular}{cc}
\includegraphics[height=0.35\textwidth]{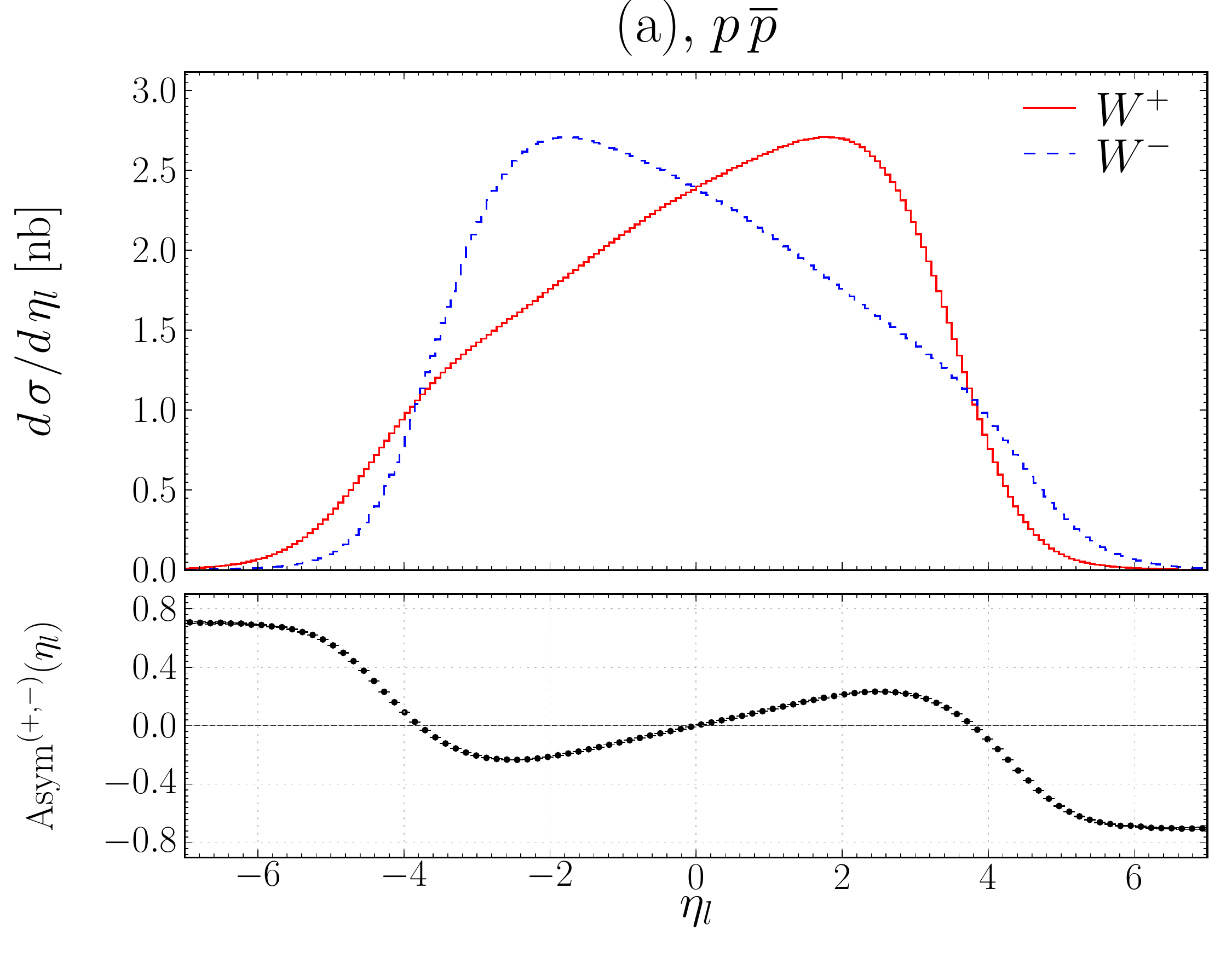} &
\includegraphics[height=0.35\textwidth]{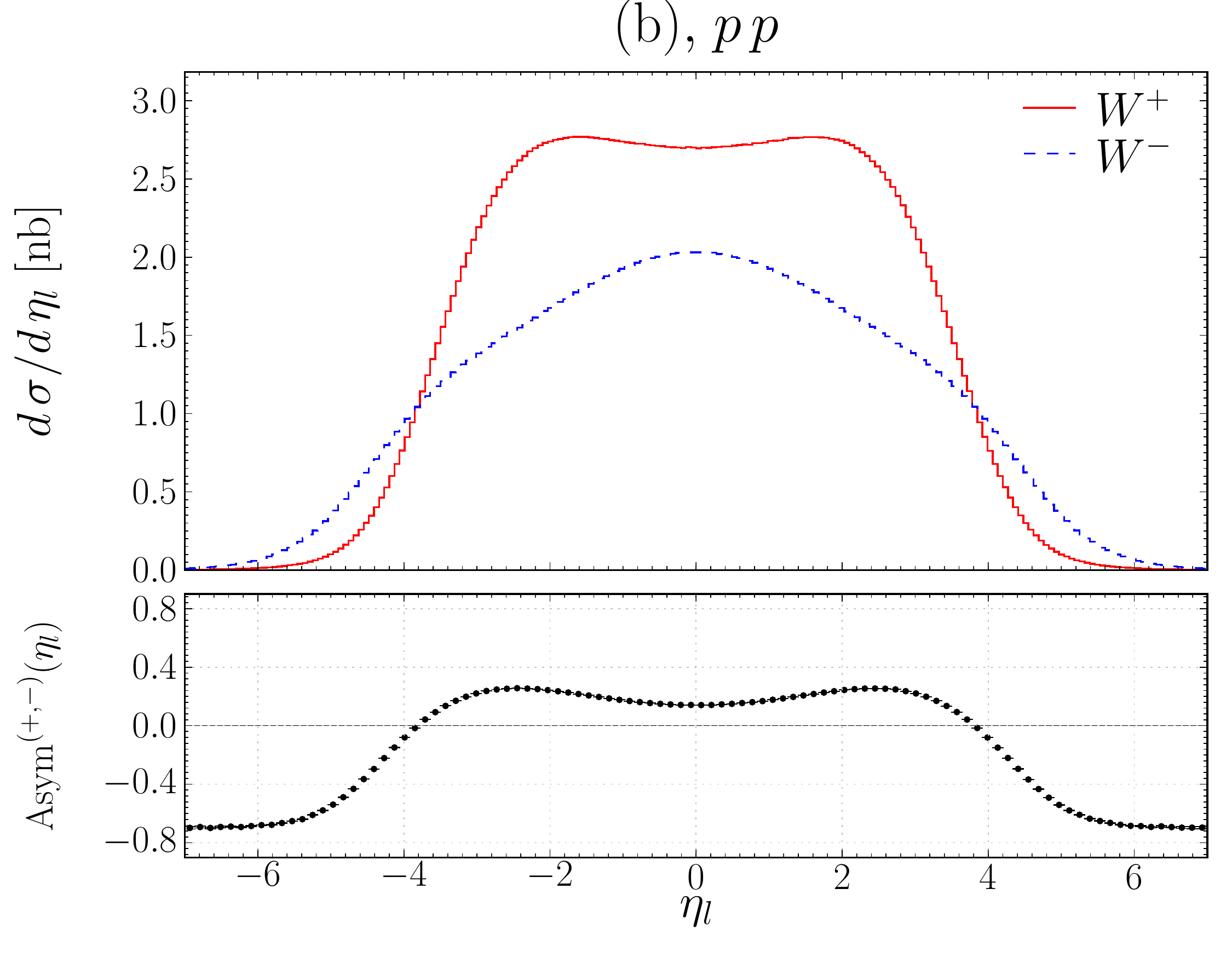} \\
\includegraphics[height=0.35\textwidth]{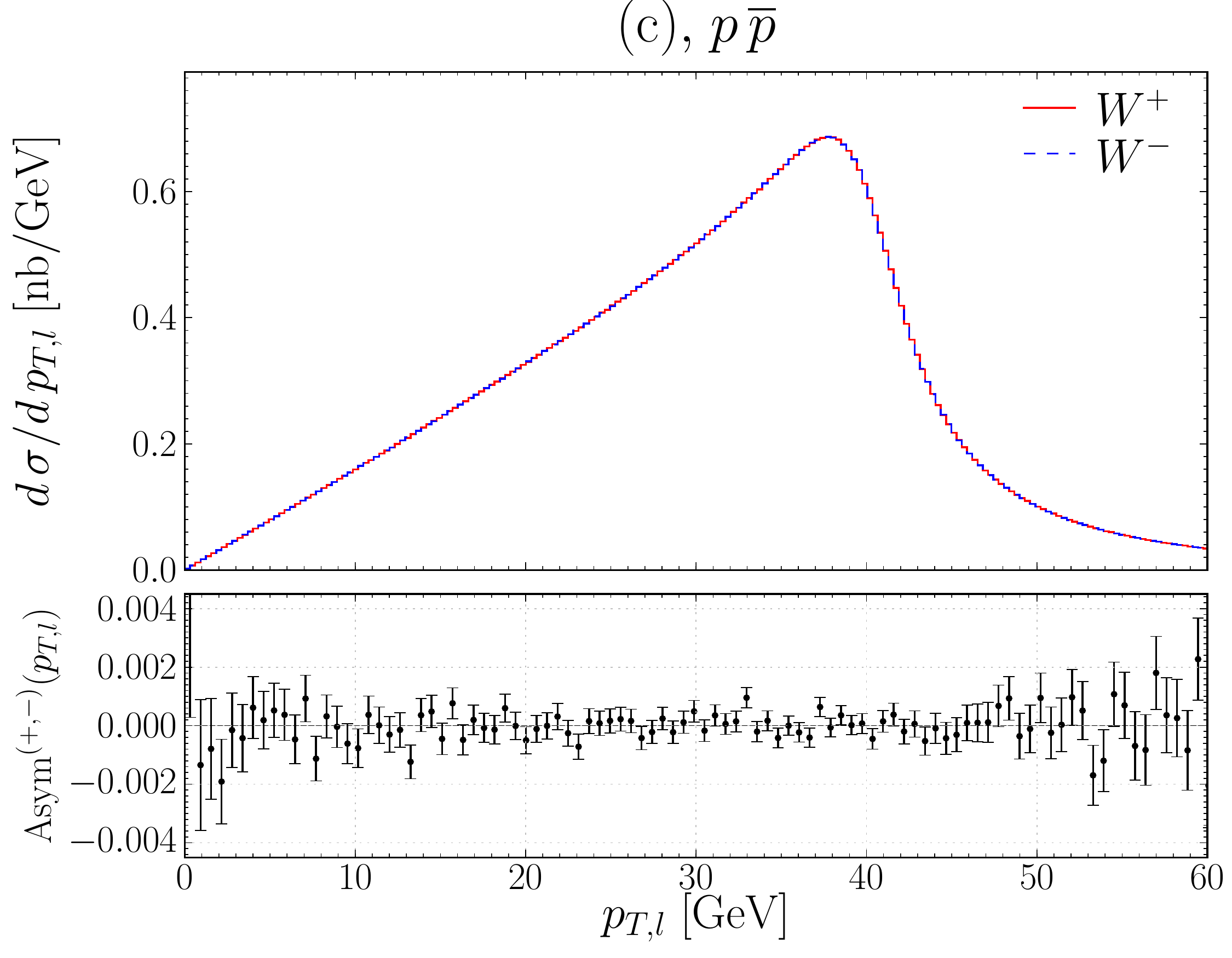} &
\includegraphics[height=0.35\textwidth]{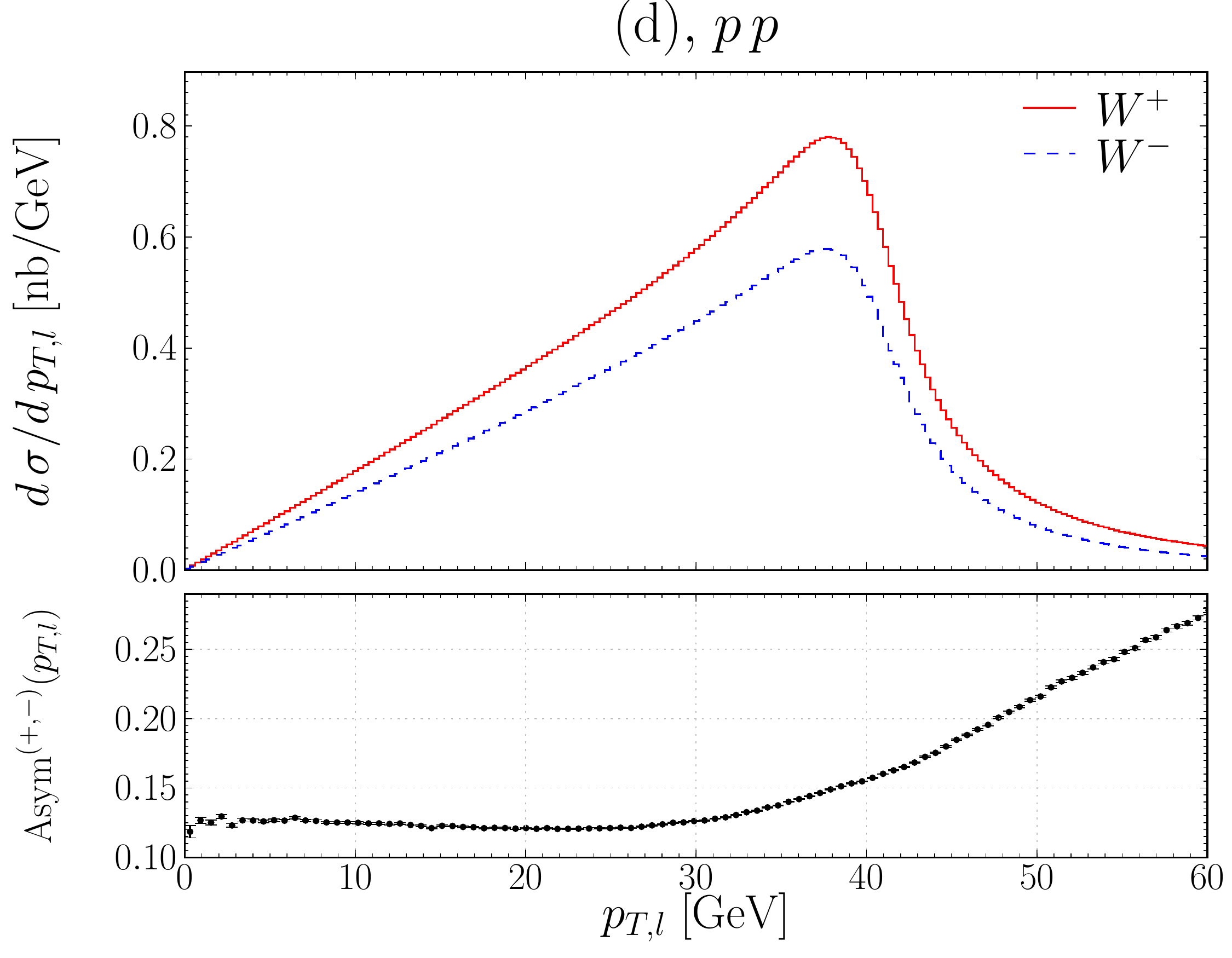} \\
\end{tabular}
\end{center}
\caption{Pseudorapidity $\eta$ and the transverse momentum
\pT\ of charged leptons from the decay of \Wpm 's, 
in \ppbar\ collisions (left panels)
and in \pp\ collisons (right panels).}
\label{WleptonyandpT}
\end{figure}

Figure~\ref{WyandpT}, taken from Ref.~\cite{Fayette:2008wt}, 
illustrates the rapidity $y$ and the
$p_{\rm T}$ of W$^{\pm}$ production, and Fig.~\ref{WleptonyandpT},
also taken from Ref.~\cite{Fayette:2008wt},
shows the pseudorapidity and the $p_{\rm T}$ for the respective decay leptons. 
The difference between the characteristics of W production and decay in \ppbar\ 
collisions and in \pp\ collisions is rather striking.
In \pp\ collisions,  the difference between \Wp\ and \Wm\ production 
is smallest at $|y| \sim 0$, because 
in this region the contribution from the annihilation of sea quarks
with sea quarks is largest. 

A common analysis of charged leptons from \Wp\ and \Wm\  
is equivalent to a W decay with equal V$-$A and V$+$A 
amplitudes, which is  parity-conserving and resembles closely
the nearly parity-conserving Z decay.
Therefore, at the Tevatron it is straightforward and justified to
calibrate in a `charge-blind' analysis the
\pT\ spectrum of charged leptons from W decay
with the \pT\ spectrum of charged leptons from Z decay.
The ensuing systematic error of the W mass at the Tevatron
is not dominant and comparable with the statistical 
error.

At the LHC, for the preponderance of \Wp\ over \Wm\ production in
\pp\ collisions, there is no cancellation at work that lends itself to
a charge-blind analysis.
If there were only sea quarks involved in the production 
of W's in \pp\ collisions, symmetry between \Wp\  and \Wm\ production
would not be broken. In practice, symmetry is broken by valence
quarks, more specifically by the difference of the
u$_{\rm v}$ and d$_{\rm v}$ PDFs of the
proton. The charge-blind analysis 
that is valid for \ppbar\ collisions at the Tevatron, 
cannot be used as template for the analysis of \pp\  collisions 
at the LHC. 

\begin{figure}[h] 
  \begin{center}
   \includegraphics[width=0.495\tw]{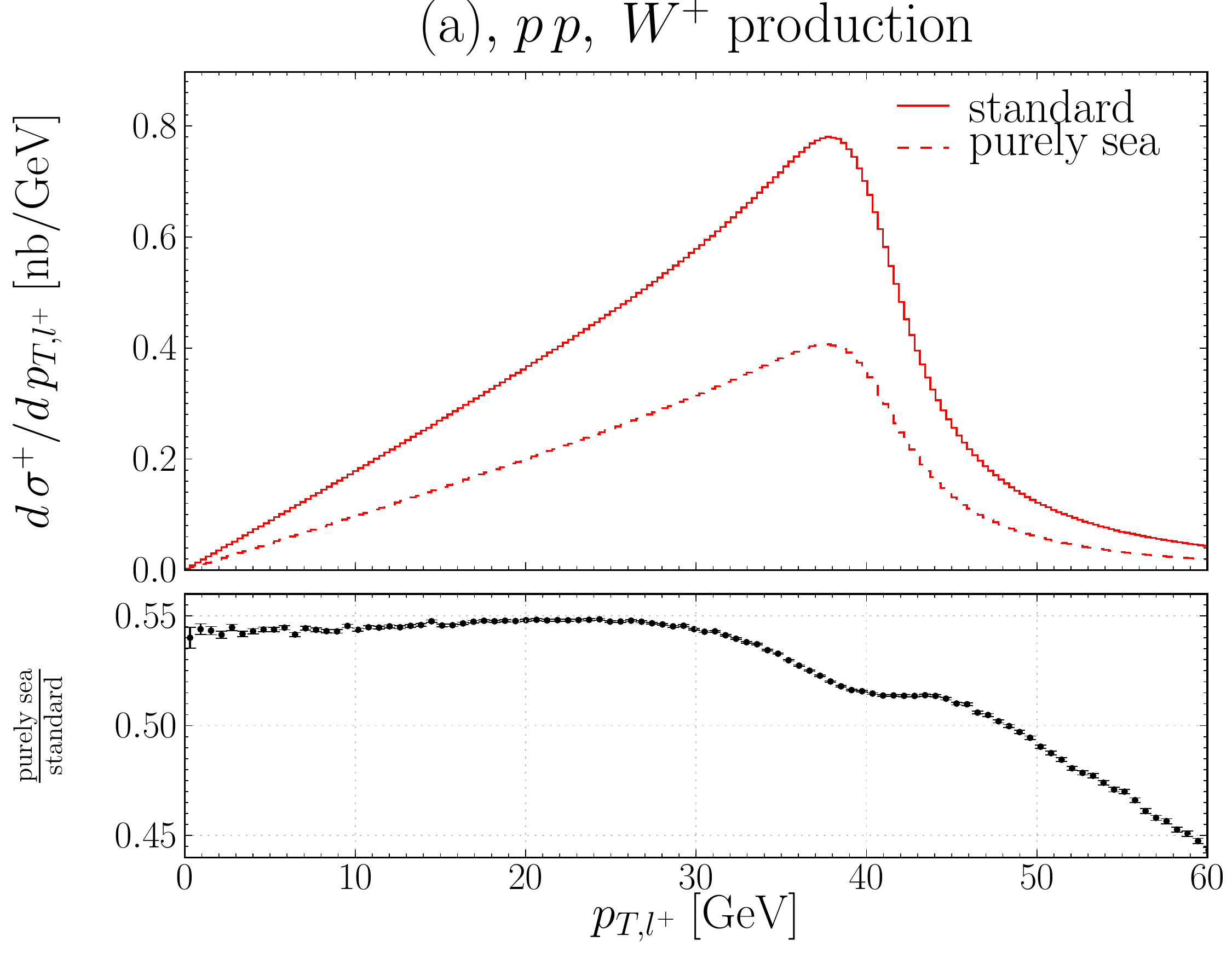}
    \hfill
   \includegraphics[width=0.495\tw]{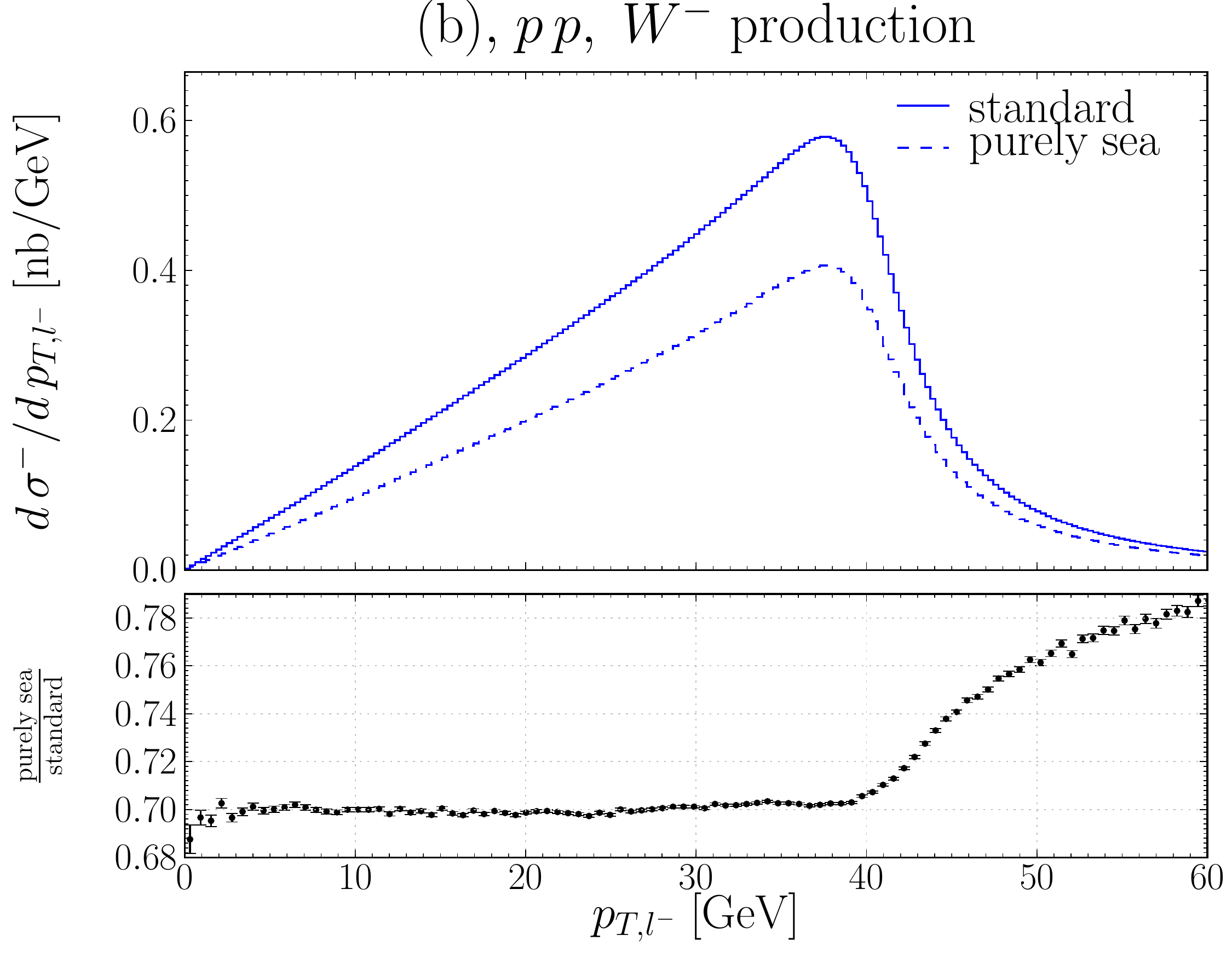}
   \caption{The effect of the valence quarks on the \pTl\ spectra of \Wp\ (left panel)
           and \Wm\ (right panel) in \pp\ collisions at the LHC.}                 
  \label{plot3}
  \end{center}
\end{figure}

The effect of the valence quarks on the \pTl\ spectra of \Wp\ and \Wm\
is shown in Fig.~\ref{plot3}.
It follows that this difference, as well as the amount
of sea quarks with u and d flavour, 
must be known with better precision than needed at the Tevatron. 
This is the LHC-specific effect from the 1st quark family. 

\begin{figure}[h]
\begin{center}
\includegraphics[width=0.90\textwidth]{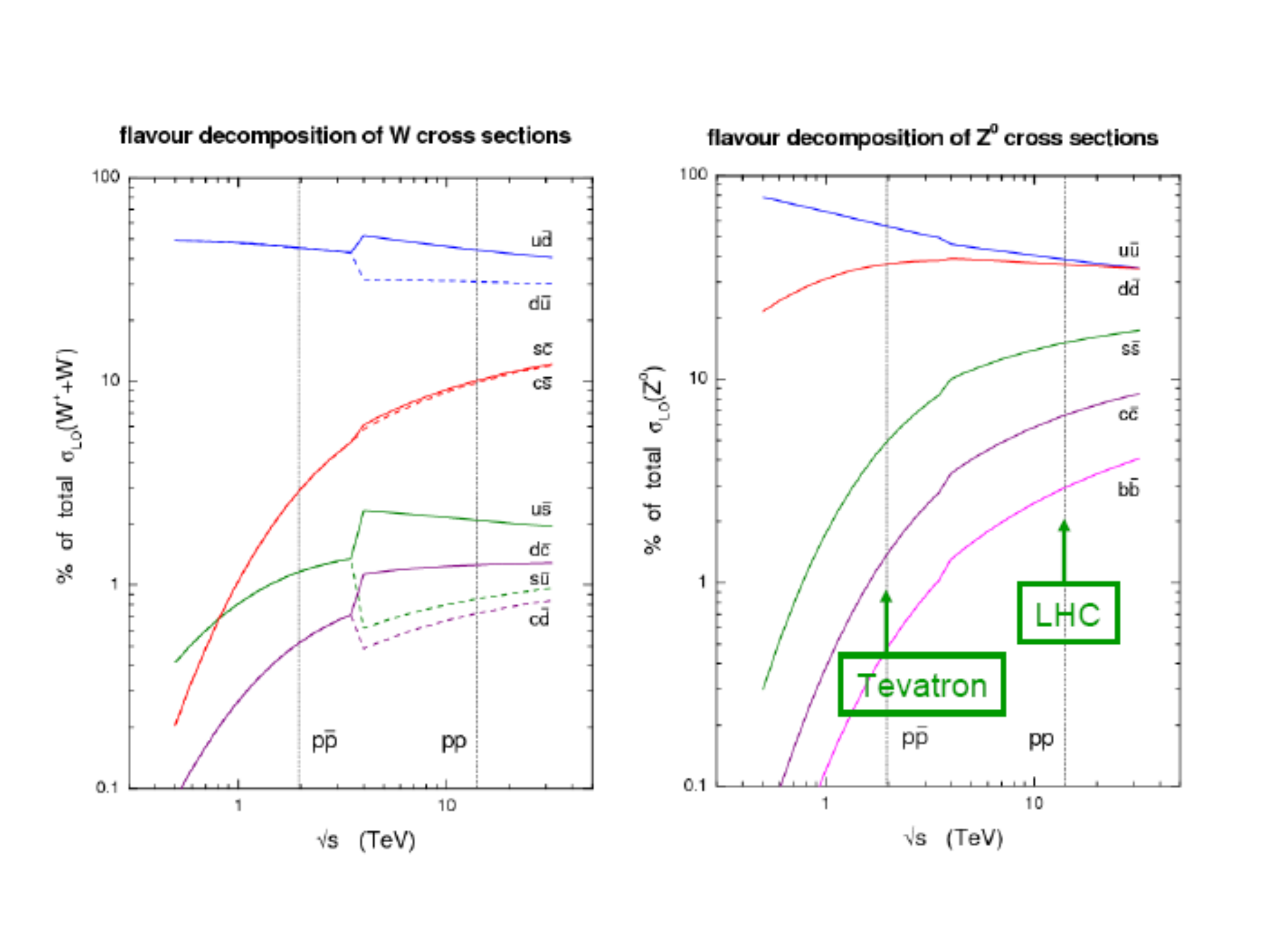}
\end{center}
\caption{Contributions of different quark--antiquark annihilations to
\Wpm\ (left panel) and 
Z (right panel) production, as a function of the beam energy.}
\label{energydependence}
\end{figure}

In Fig.~\ref{energydependence},
taken from Ref.~\cite{Stirling},
the contributions of different quark--antiquark annihilations
to the \Wp , \Wm\ and Z cross-sections
are shown as a function of beam energy, and specifically for 
the Tevatron and LHC  energies. The much stronger contributions from
c and b quarks at the LHC energy are noteworthy.  

\begin{figure}[h] 
  \begin{center}
    \includegraphics[width=0.495\tw]{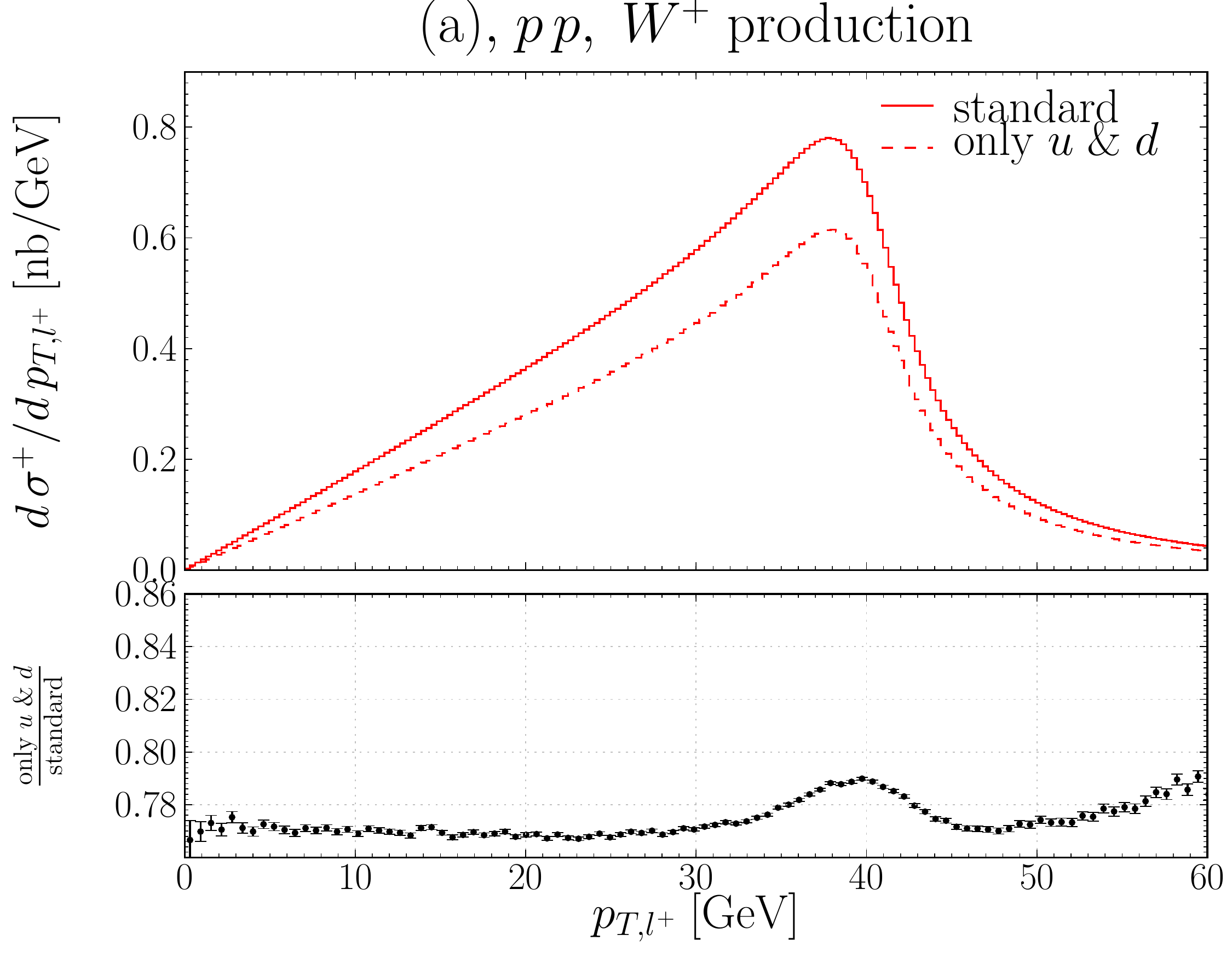}
    \hfill
    \includegraphics[width=0.495\tw]{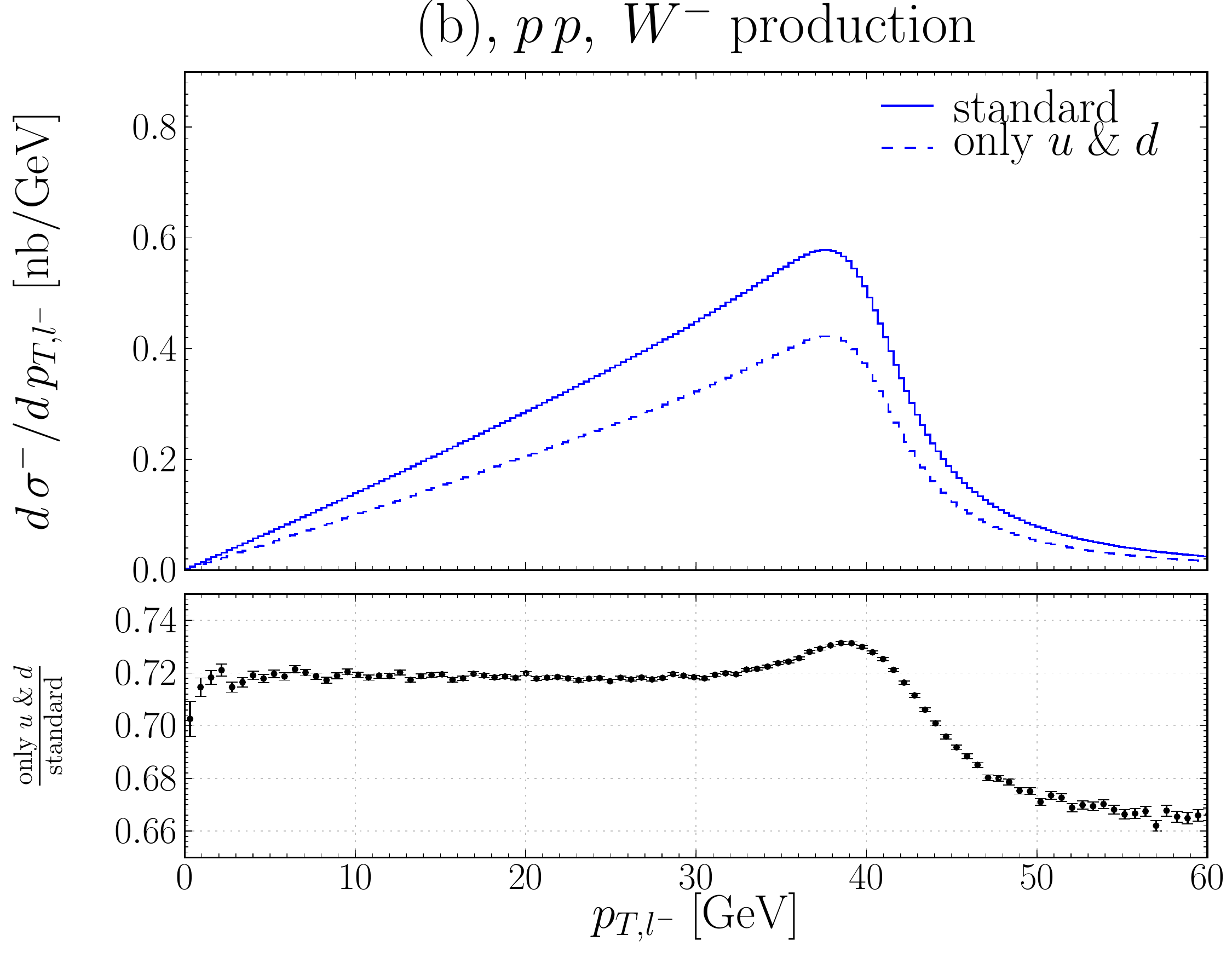}
    \caption{The effects of s and c quarks on the \pTl\ spectra of \Wp\ (left panel)
           and \Wm\ (right panel) in \pp\ collisions at the LHC.}               
    \label{plot5}
  \end{center}
\end{figure}

The effects of s and c quarks on the \pTl\ spectra of \Wp\ (left panel)
and \Wm\ (right panel) in \pp\ collisions at the LHC are shown in Fig.~\ref{plot5}.
The partner of a c quark to form a 
W boson, the s quark, has different distributions in $x$ 
and \kT\ than a c quark that is needed as partner 
to form a Z boson. Therefore, to know the difference between the
s and c PDFs of the proton is important. 
This is the LHC-specific effect of the 2nd quark family. 
 
Although both for W and Z production the contribution from b quarks is  
negligible at the Tevatron, 
and also for W production at the LHC, 
the contribution of b quarks to  
Z production is important at the LHC and a sufficiently precise knowledge of
the PDF of the b quark is needed.
This is the LHC-specific effect of the 3rd quark family. 

The importance of the LHC-specific 
intricacies of the production and decay 
mechanisms, and
their effect on the \pT\
spectra of decay leptons of \Wp , \Wm\ and Z, has 
been missed in the LHC physics studies made so
far. Not a single study 
made a difference 
between charged leptons from \Wp\  and \Wm\ decays. As a consequence
of these shortcuts, 
unrealistically small errors 
at or below the 10~\MeVcsq\ level 
were reported for the W mass measurement at the LHC\footnote{The discussion in this paper applies {\em mutatis mutandis\/} also to the determination of the W mass from
$m_{\rm T}$ spectra. The determination of $m_{\rm T}$ involves the
reconstruction of the neutrino transverse momentum as missing
transverse momentum. The systematic error of this measurement which involves 
the reconstruction of the hadronic system, is too large to
be useful for the measurement of the W mass at the 10~\MeVcsq\ level, {\it inter alia\/} 
for reasons of reconstruction efficiency and acceptance close to the beam pipe.}

\subsection{A biased W mass}
\label{BIASEDWMASS}

In this section, it is argued that the current precision of pertinent proton PDFs is
overestimated. It is shown that with realistic errors of these PDFs, and with
correlations taken into account, 
the advocated W mass precision of 10~\MeVcsq\ or 
better~\cite{mW-ATLAS, mW-CMS} is much too optimistic. 
If the current underestimation of systematic uncertainties is not rectified
the W mass from the LHC may be seriously biased.

\begin{figure}[h]
\begin{center}
\includegraphics[width=0.6\textwidth]{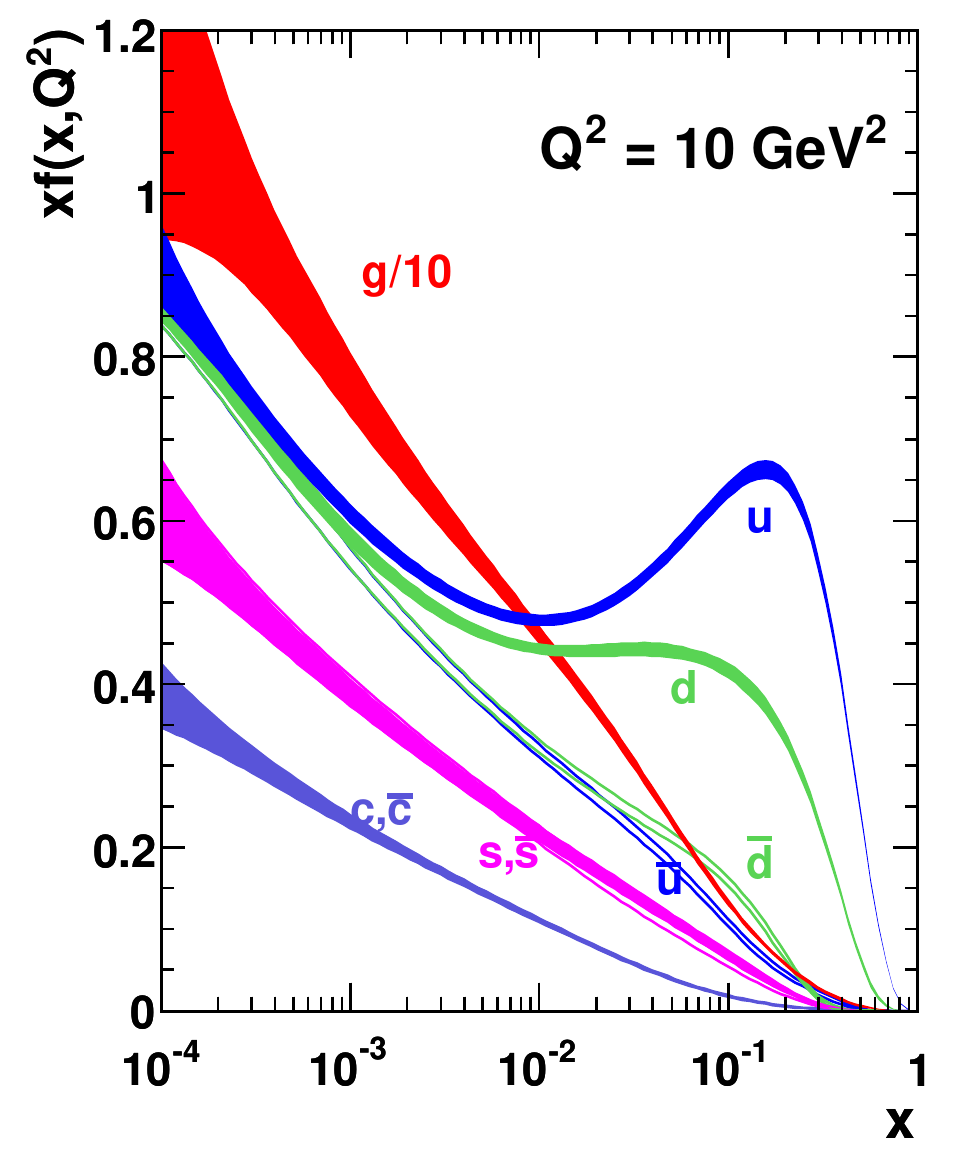}
\end{center}
\caption{The MSTW--2008 proton PDFs  
at $Q^2 = 10$~GeV$^2$/{\it c}$^4$; the widths of the bands characterize the estimated
uncertainty.}
\label{protonpdfs}
\end{figure}

The current understanding of the parton density functions is summarized in 
Fig.~\ref{protonpdfs} which shows the 
MSTW--2008 set~\cite{Stirling}.
It is advocated and widely believed that the proton PDFs  are precise enough 
not to pose a limitation for LHC data analysis. 
For example, the $u_{\rm v}$ and
$d_{\rm v}$ PDFs are claimed to be precise to
2\%~\cite{Stirling}. 

\begin{figure}[h]
\begin{center}
\begin{tabular}{cc}
\includegraphics[width=0.4\textwidth]{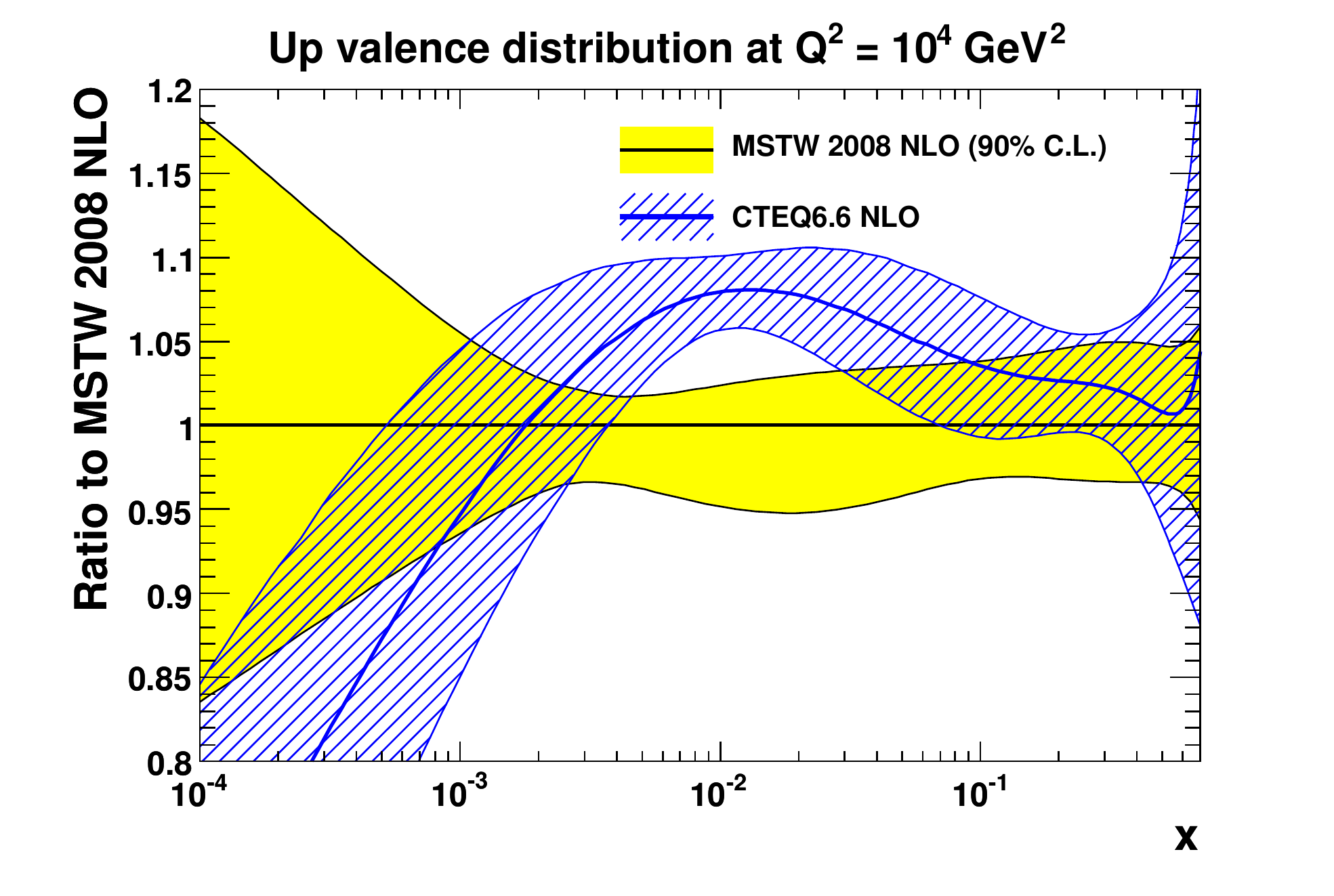} & 
\includegraphics[width=0.4\textwidth]{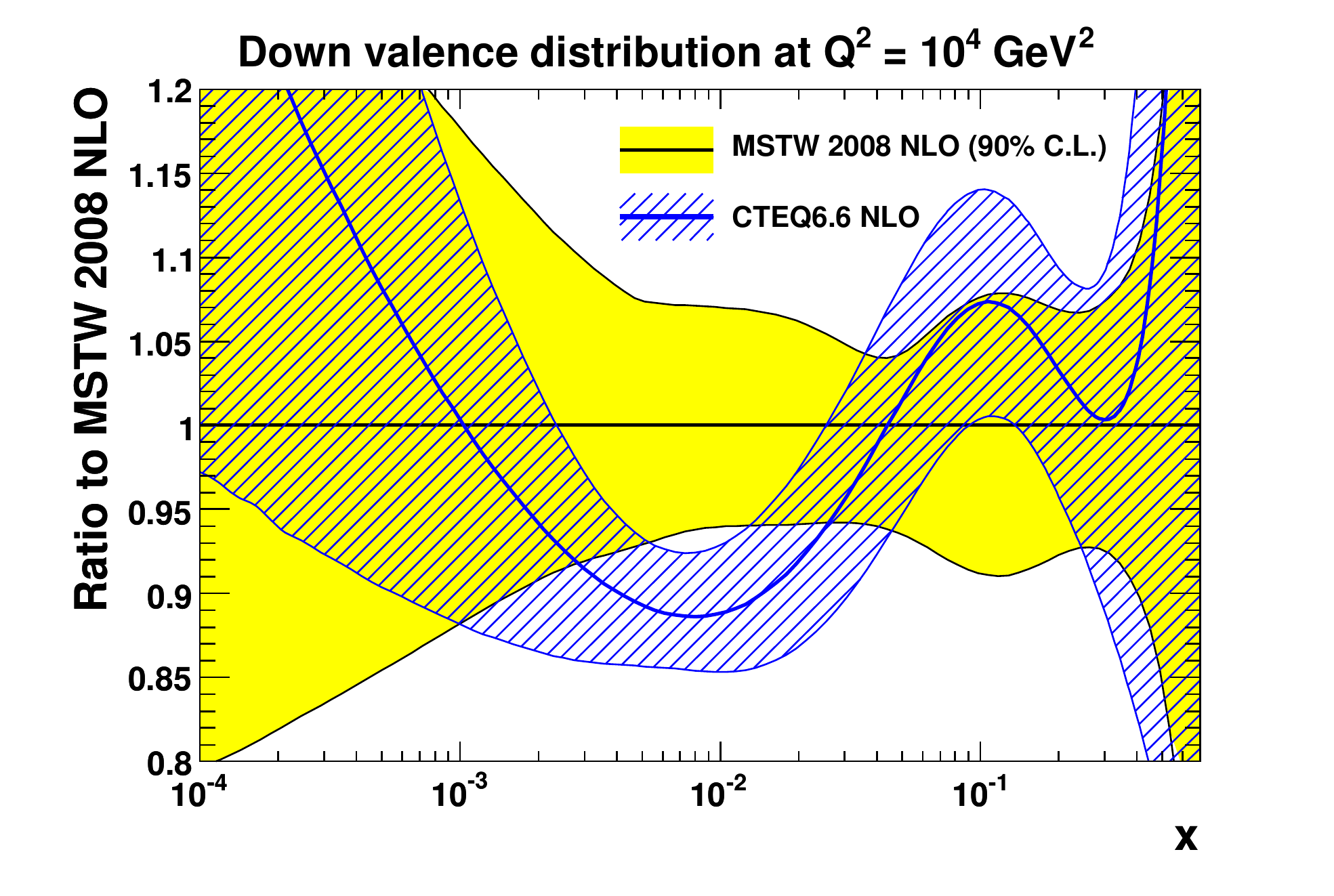} \\
\includegraphics[width=0.4\textwidth]{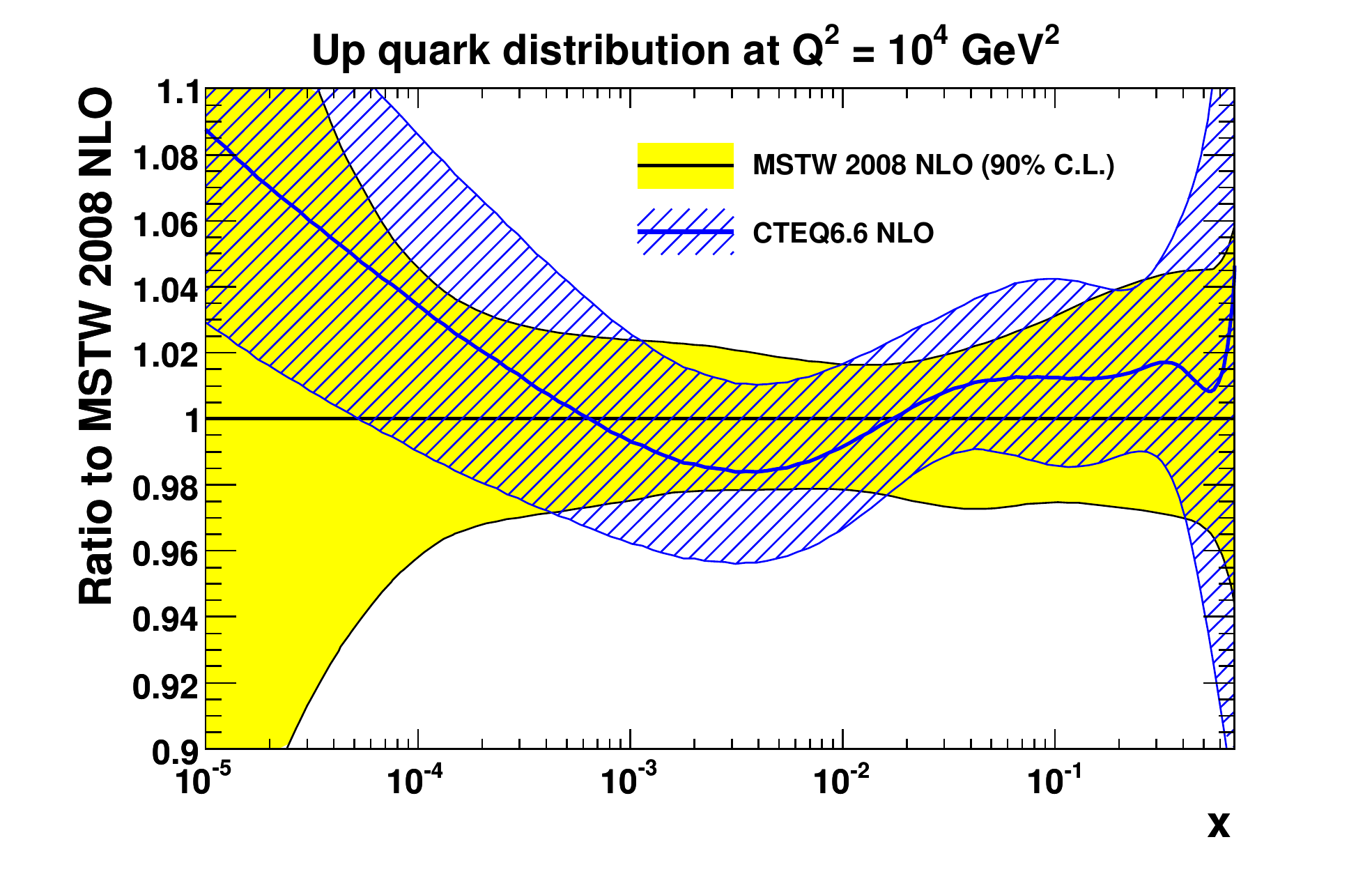} &
\includegraphics[width=0.4\textwidth]{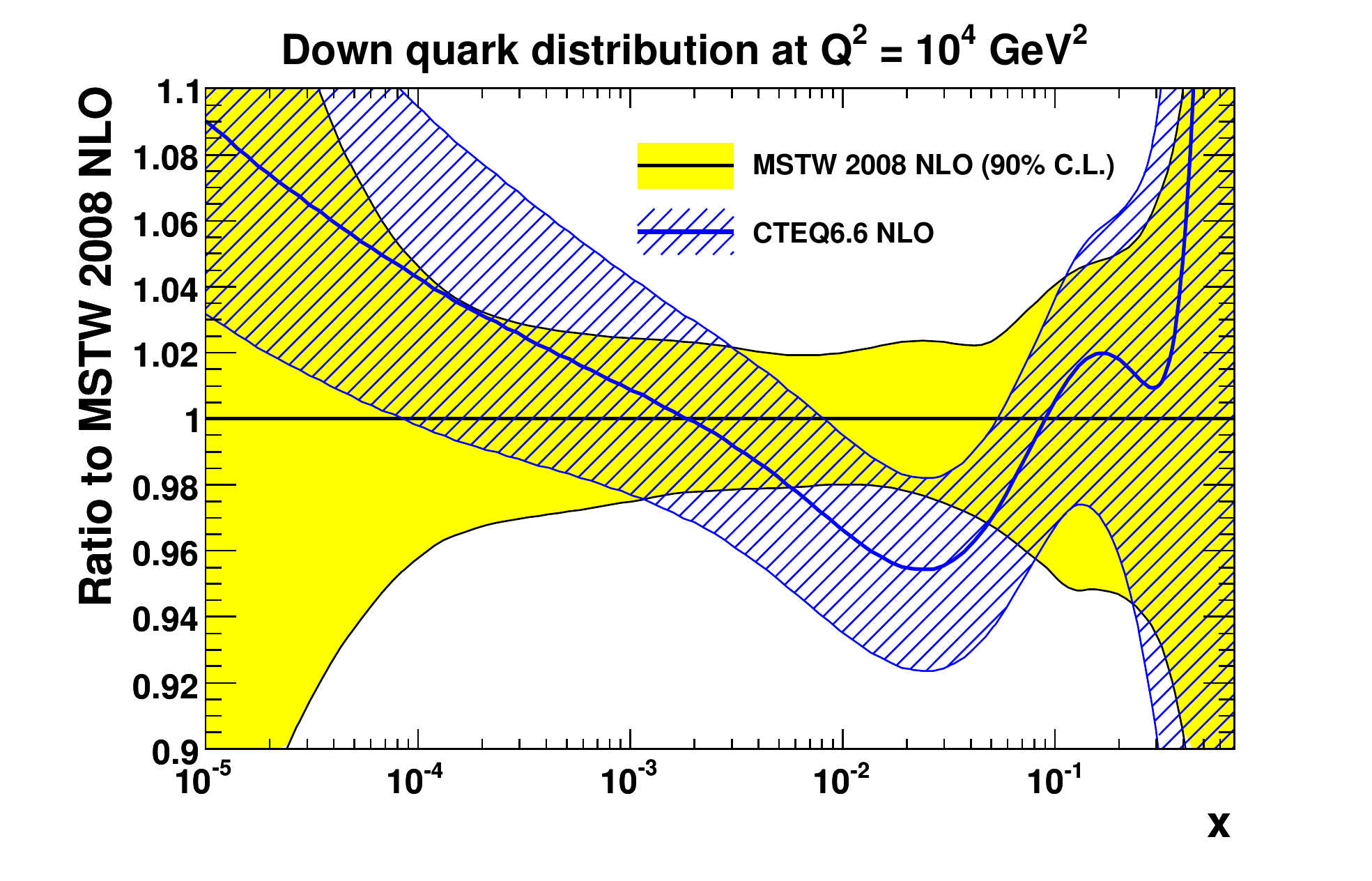} \\
\includegraphics[width=0.4\textwidth]{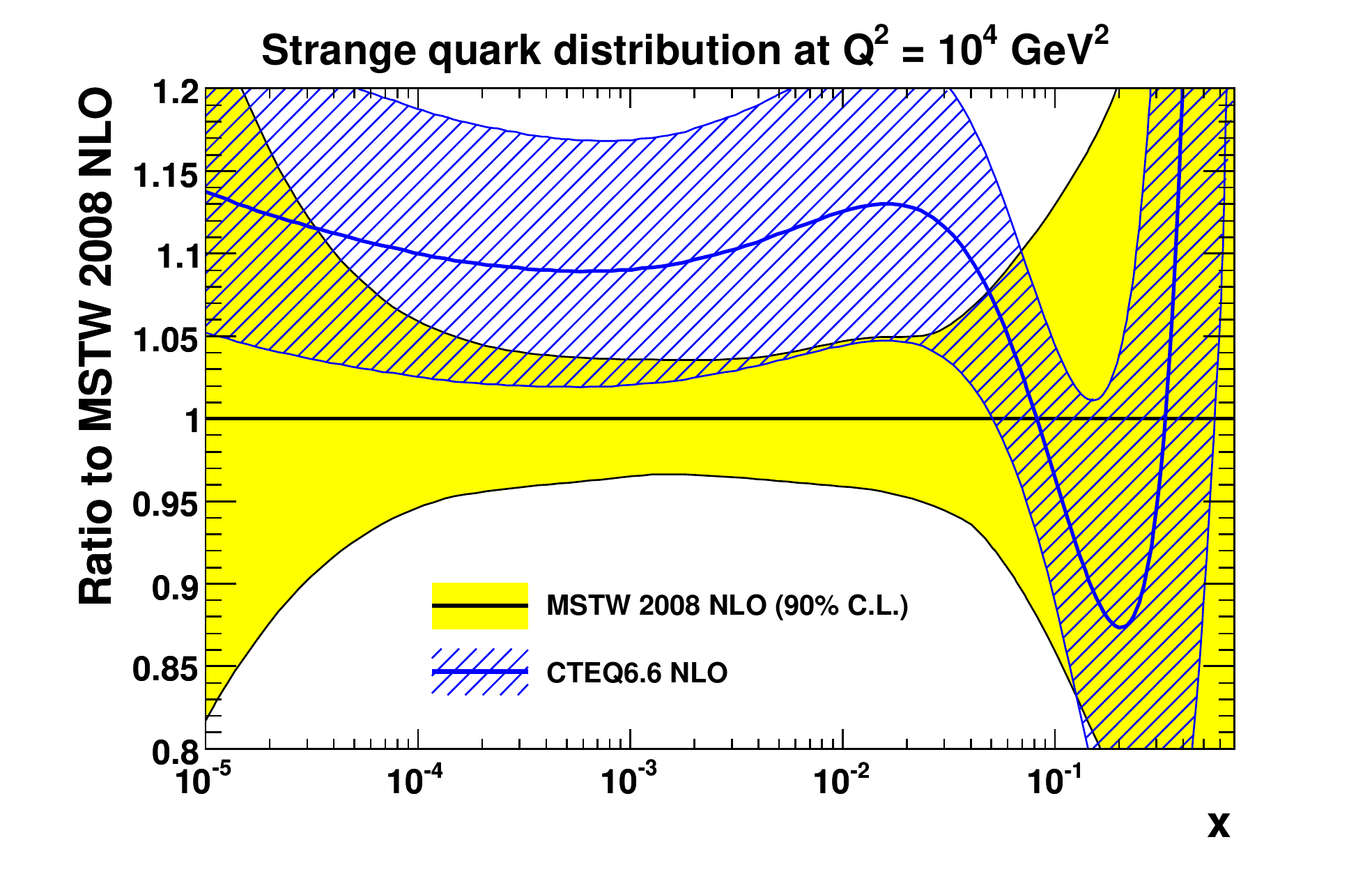} &
\includegraphics[width=0.4\textwidth]{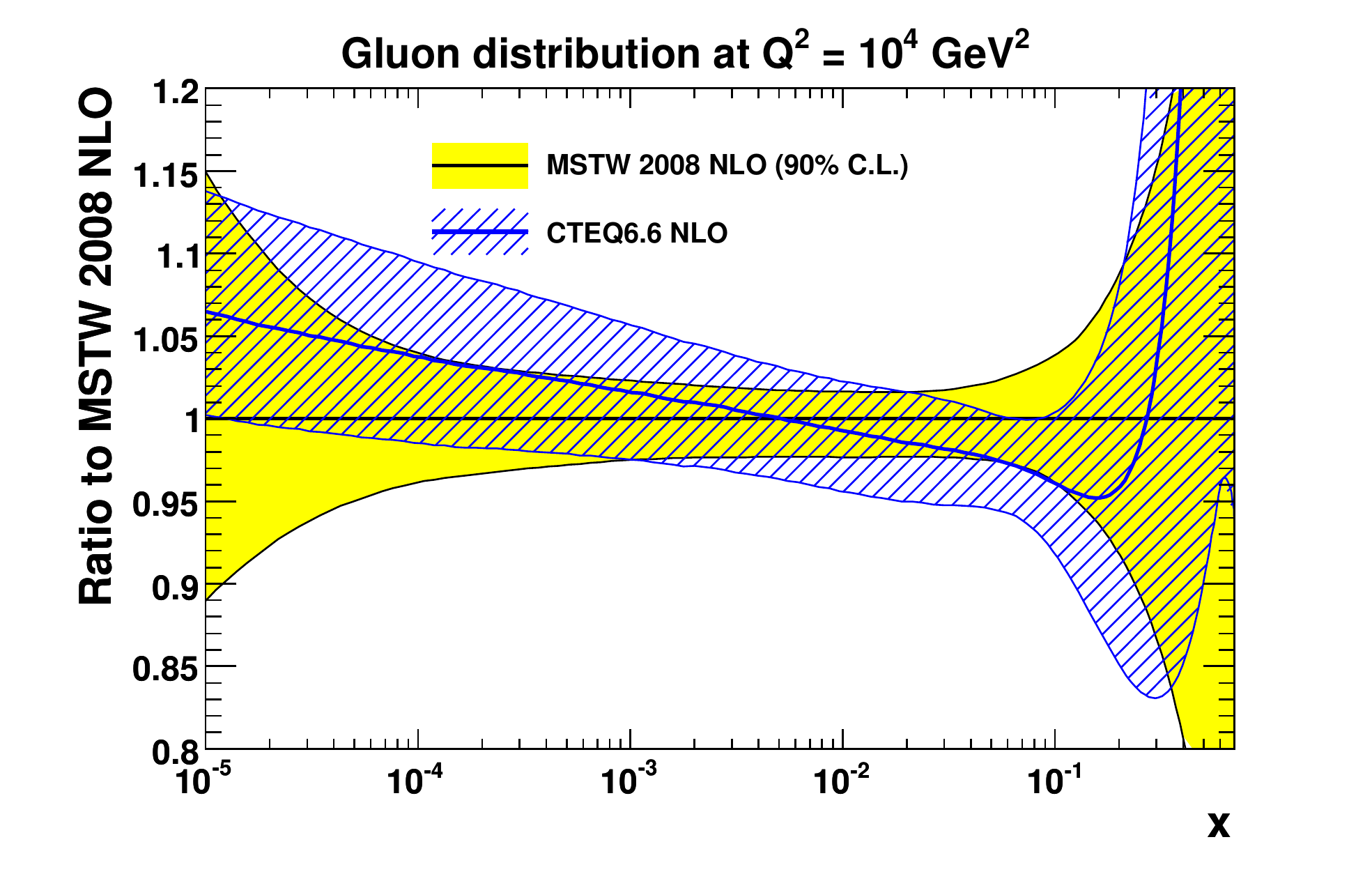} \\
\end{tabular}
\end{center}
\caption{Comparison of the CTEQ6.6 and MSTW2008 (NLO) PDFs
of u$_{\rm v}$, d$_{\rm v}$, u, d and s quarks, 
and of gluons.}
\label{CTEQvsMSTW}
\end{figure}

Why then differ the CTEQ~\cite{CTEQ} and MSTW~\cite{MSTW} 
proton PDFs by much more than 2\%, 
as shown in Fig.~\ref{CTEQvsMSTW} taken from Ref.~\cite{MSTW}, 
although they stem largely from the same input data? 
A 5\% error of the PDFs of the u and d quarks 
appears more realistic. 

\begin{figure}[h]
\begin{center}
\includegraphics[width=0.60\textwidth]{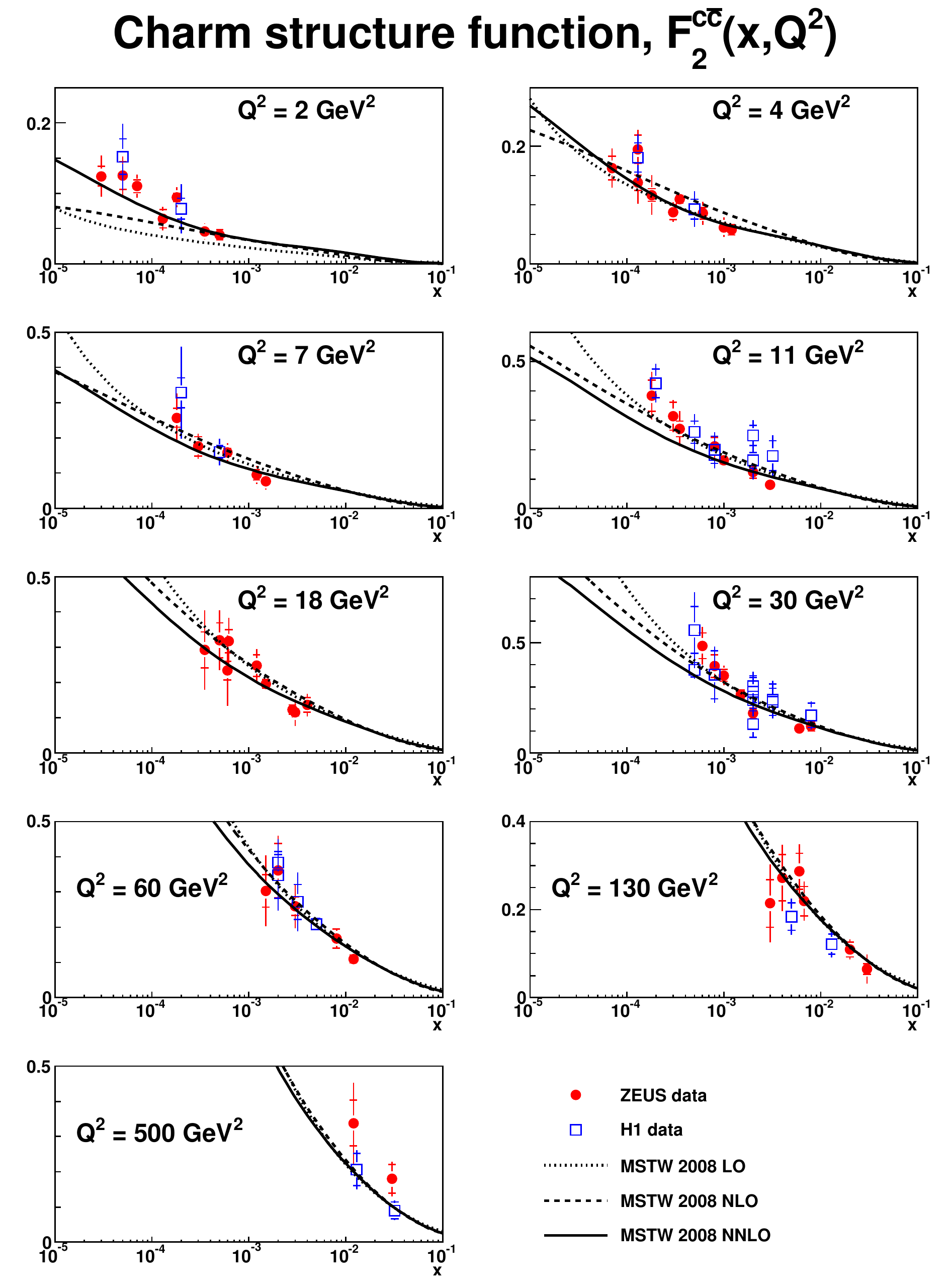} 
\end{center}
\caption{The measured PDF of the c quark, and the MSTW2008 fit,  
at different values of $Q^2$.}
\label{MSTWcquark}
\end{figure}

\begin{figure}[h]
\begin{center}
\includegraphics[width=0.60\textwidth]{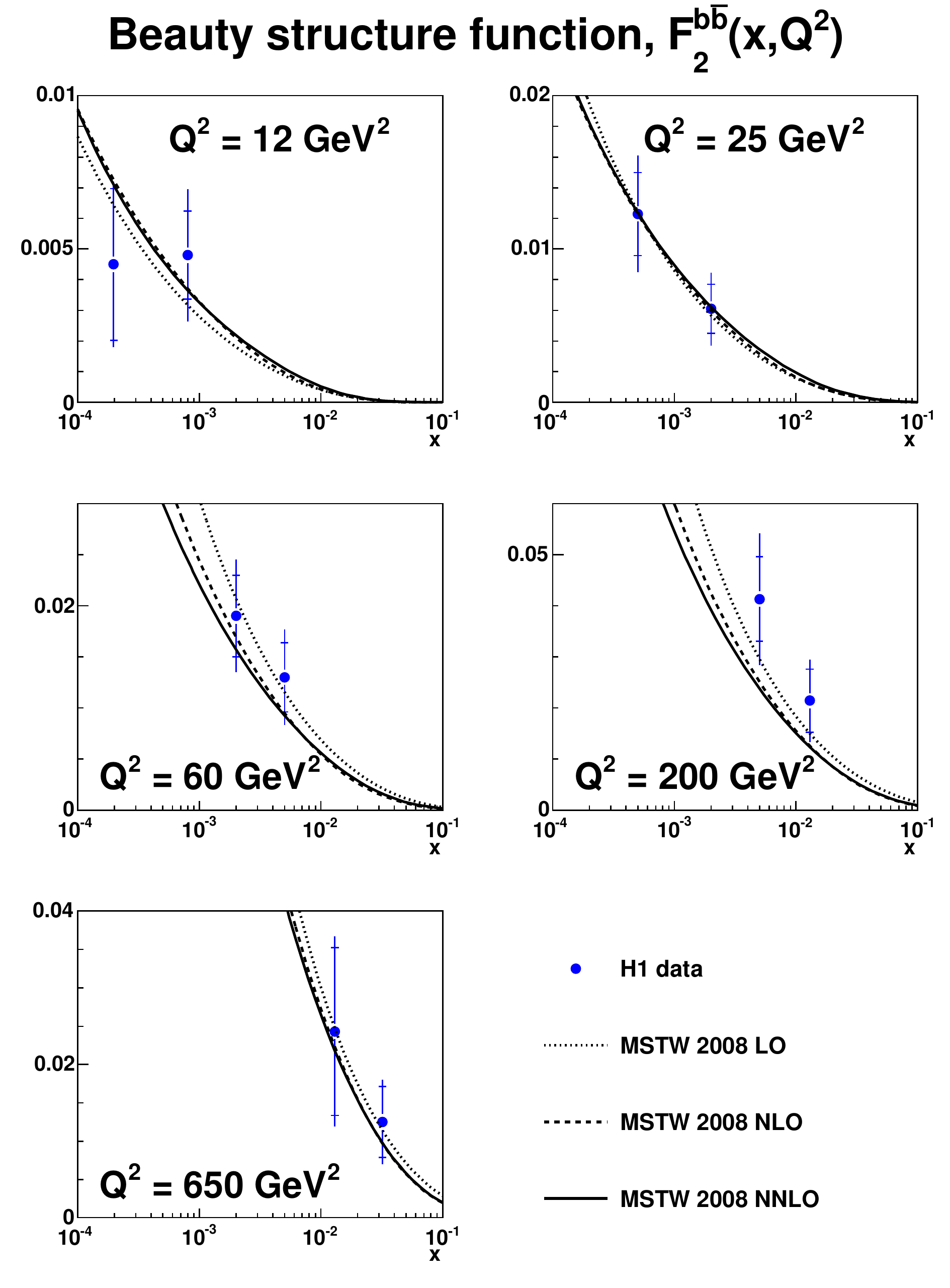} 
\end{center}
\caption{The measured PDF of the b quark, and the MSTW2008 fit, 
at different values of $Q^2$.}
\label{MSTWbquark}
\end{figure}

The present experimental uncertainty of the PDF of the c quark is at the 10\%
level\footnote{Theoretical calculations of heavy-quark PDFs from the 
gluon PDF are claimed to have a smaller error margin.} , see Fig.~\ref{MSTWcquark} taken from Ref.~\cite{MSTW}.
The present experimental uncertainty of the PDF of the b quark
is at the 20\%
level, see Fig.~\ref{MSTWbquark} taken from Ref.~\cite{MSTW}.

Another problem for the use of current proton PDFs in the analysis
of W and Z production and decay at the LHC arises from
`compensating' PDF changes:
a change of the PDF of one quark can be compensated by a change 
of the PDF of the other quark of the same family that 
leaves the Z rapidity distribution nearly invariant
and hence escapes detection\footnote{The condition of invariance of the Z rapidity distribution,
and hence invisibility even in high-statistics data samples,
is decisive: if the measured Z rapidity distribution looked differently 
than expected from the current proton PDFs, 
an appropriate change of the proton PDFs would be unavoidable.}.

For the 3rd quark family compensating PDF changes 
are obviously not possible.

The above uncertainties of PDFs are incorporated in
the simulation of \pT\ spectra from \Wp , \Wm\ and Z leptonic decays.
This simulation uses  
the LHAPDF package~\cite{LHAPDF} of PDFs, and 
PYTHIA 6.4~\cite{PYTHIA} for the modelling of the QCD/QED initial-state
parton shower and its hadronization; the
transverse momentum \kT\ of quarks and antiquarks is the
one incorporated in PYTHIA.
The tool for event generation is WINHAC 1.31~\cite{WINHAC},
a Monte Carlo generator for single W production in hadronic
collisions, and subsequent leptonic decay. WINHAC includes also neutral-current
processes with $\gamma$ and Z bosons in the intermediate state.
The novel feature of WINHAC is that it
describes W and Z production and decay in 
terms of spin amplitudes~\cite{PlaczekJadach}. These 
involve, besides all possible spin configurations of the W and Z bosons,
also the ones of the initial- and final-state fermions. 
The advantage of this approach is that one has explicit control 
over all spin states, and thus over transverse and longitudinal 
boson polarization amplitudes
and their interferences.
 
As an example LHC detector, ATLAS is chosen. Charged leptons from
W and Z decays are accepted with $p_{\rm T} > 20$ GeV/{\it c}
and $|\eta| < 2.5$. The event statistics  correspond to an
integrated luminosity of 10~fb$^{-1}$.
Both the electron- and muon decay channels of W and Z
are considered. 
Since in \pp\ collisions the spectra of positive and 
negative leptons are to be analyzed separately, it is
natural to make the same distinction also for the leptons
from Z decay. Along this line of reasoning, 
`Z$^+$' and `Z$^-$' lepton $p_{\rm T}$ spectra are generated, 
in analogy to 
`W$^+$' and `W$^-$' lepton $p_{\rm T}$ spectra\footnote{This appears 
appropriate as 
a non-zero longitudinal Z polarization causes 
the $p_{\rm T}$ spectra of the 
positive and negative decay leptons to be slightly different,
for the charge-dependent correlation of the Z spin 
with the emission of charged decay leptons.}. 
All spectra are generated
with various proton PDF configurations.
The Z$^+$ and Z$^-$ lepton $p_{\rm T}$ spectra are corrected
for the evolution from $Q^2 = M^2_{\rm W}$ to 
$Q^2 = M^2_{\rm Z}$.

From a fit of the Jacobian peaks in the
$p_{\rm T}$ distributions and by calibrating with the known Z mass,
the W$^+$ and W$^-$ masses are determined.  

For technical reasons, not $M_{\rm W^+}$ and $M_{\rm W^-}$
are separately determined 
but, equivalently, the average $(M_{\rm W^+} + M_{\rm W^-})/2 = \MW$ and 
the difference $M_{\rm W^+} - M_{\rm W^-}$ of the masses.

Table~\ref{WMbias1} lists the biases of 
\MW\ and
of $M_{\rm W^+} - M_{\rm W^-}$ caused by compensating
changes of the PDFs of quarks of the 1st family\footnote{The 
differences $M_{\rm W^+} - M_{\rm W^-}$ are taken from  
Ref.~\cite{Fayette:2008wt}.}. If the lepton rapidity range is reduced from 
$|\eta| < 2.5$ to $|\eta| < 0.3$, the quoted errors would reduce roughly by
a factor of two.
\begin{table}[h]
\begin{center}
\begin{tabular}{|c|c|c|}
\hline
   & $\Delta \MW $ 
   & $\Delta [(M_{\rm W^+} - M_{\rm W^-})]$ \\
\hline
\hline
$u_{\rm v}^{\rm bias}$ = $1.05 \,u_{\rm v}$ & $+79$~MeV & $+115$~MeV \\
$d_{\rm v}^{\rm bias}$ = $d_{\rm v} - 0.05 \,u_{\rm v}$ & & \\
\hline
$u_{\rm v}^{\rm bias}$ = $0.95 \,u_{\rm v}$ & $-64$~MeV & $-139$~MeV \\
$d_{\rm v}^{\rm bias}$ = $d_{\rm v} + 0.05 \,u_{\rm v}$ & & \\
\hline  
\end{tabular}
\vspace*{2mm}
\caption{Biases from uncertainties in the 1st quark family.}
\label{WMbias1}
\end{center}
\end{table}

Table~\ref{WMbias2}
lists the biases of \MW\  and
of $M_{\rm W^+} - M_{\rm W^-}$ caused by compensating
changes of the PDFs of quarks of the 2nd family.
\begin{table}[h]
\begin{center}
\begin{tabular}{|c|c|c|}
\hline
   & $\Delta \MW $ 
   & $\Delta [(M_{\rm W^+} - M_{\rm W^-})]$ \\
\hline
\hline
$c^{\rm bias}$ = $0.9 \,c$ & $+148$~\MeVcsq\ & $+17$~\MeVcsq\ \\
$s^{\rm bias}$ = $s + 0.1 \,c$  & & \\
\hline
$c^{\rm bias}$ = $1.1 \,c$ & $-111$~\MeVcsq\ & $-11$~\MeVcsq\ \\
$s^{\rm bias}$ = $s - 0.1 \,c$  & & \\
\hline  
\end{tabular}
\vspace*{2mm}
\caption{Biases from uncertainties in the 2nd quark family.}
\label{WMbias2}
\end{center}
\end{table}

Table~\ref{WMbias3}
lists the biases of \MW\ caused by 
changes of the PDF of the b quark.

The conclusion is, when allowing for compensating PDF changes 
and a realistic PDF error margin, that there is no way to obtain \MW\ with a 
precision at the 10~\MeVcsq\ level with the
currently available proton PDFs.

Can the pertinent PDFs be improved with data from ongoing lepton--nucleon
scattering experiments, from the Tevatron, or from the LHC? As will be discussed 
in Section~\ref{WAYSFORWARD}, the answer is no. New avenues 
of experimentation are asked for.

\section{An LHC-specific programme}
\label{LHCSPECIFICPROGRAMME}

A charge-blind analysis like for \ppbar\ collisions at the Tevatron is not appropriate
for \pp\ collisions at the LHC. But there are more good reasons to think of a generic,
LHC-specific, measurement and analysis programme:

\begin{enumerate}

\item The measurement of electroweak parameters should be based on a set of
observables with reduced sensitivity to systematic measurement errors and to
theoretical uncertainties of perturbative and non-perturbative QCD.

\item The two-dimensional PDFs in the W and Z analysis at the LHC should
be defined such that compatibility is maintained with the 
\kT -integrated PDFs used at the 
Tevatron, and with the missing high-precision \kT -integrated PDFs that are 
to come from new experimental avenues. 
 
\item To avoid theoretical uncertainties in the 
modelling of the lepton \pT\ distributions in leptonic W and Z decays,
$Q^2$-scale effects should be determined from the data.

\item To calibrate the lepton momentum with sufficient precision,
special data-taking actions should be undertaken.

\end{enumerate}

An LHC-specific measurement and analysis programme along these lines is outlined below.

\subsection{Inclusive cross-sections of charged leptons from W and Z decay}
\label{CROSS-SECTIONS}

At hadron colliders, the best precision of electroweak parameters is expected from leptons from purely leptonic decays of W and Z. Since the kinematical variables of neutrinos can only be inferred from measurements involving hadrons and hence are subject to larger measurement uncertainties, only observables based on charged 
leptons \lpm\ (more specifically: electrons and muons) are considered. 

There are three classes: events with one l$^+$, events with one l$^-$, and events with one oppositely charged lepton pair l$^+$l$^-$. It is assumed that these events result from the decays of \Wp , \Wm , and Z\footnote{Throughout this paper, Z stands
for Z/$\gamma$.}. Corrections for acceptance, trigger efficiency, resolution effects,  and losses from selection cuts have been applied\footnote{Methods of selecting event samples which minimize biases in acceptance and efficiency corrections between W and Z are discussed in Ref.~\cite{Krasny:2007cy}.}. All background is assumed to be subtracted.

The following five inclusive cross-sections are measured:
\begin{eqnarray}
\Sigma_{{\rm W}^+} (\pTl , \etal  ) & = & {\rm d}^2 \sigma / ({\rm d}\pTlp {\rm d}\etalp)  \; , 
  \\
\Sigma_{{\rm W}^-} (\pTl , \etal  ) & = & {\rm d}^2 \sigma / ({\rm d}\pTlm {\rm d}\etalm) \; ,
  \\
\Sigma_{\rm Z} (\Mll, \pTll ,  \yll) & = &
{\rm d}^3 \sigma / ({\rm d}\Mll {\rm d}\pTll {\rm d} \yll) \; ,
  \\
\Sigma_{{\rm Z}^+} (\Mll, \pTll, \yll, \pTl , \etal  ) & = &
{\rm d}^5 \sigma / ({\rm d}\Mll {\rm d}\pTll {\rm d}\yll {\rm d}\pTlp {\rm d}\etalp) \; , 
  \\
\Sigma_{{\rm Z}^-} (\Mll, \pTll, \yll, \pTl , \etal  ) & = &
{\rm d}^5 \sigma / ({\rm d}\Mll {\rm d}\pTll  {\rm d}\yll {\rm d}\pTlm {\rm d}\etalm) \; ,
\label{Sigmas}
\end{eqnarray}
where \pTlp\ and \etalp\  (\pTlm\ and  \etalm) denote the transverse momentum and 
pseudorapidity of positively (negatively) charged leptons, and \Mll , \yll\  and   \pTll\ 
the mass, rapidity and transverse momentum of the charged lepton pair. The latter two cross-sections are not independent, only one of them, or a suitable combination,  can be used. 

There is one correction of the above cross-sections that still needs to be applied, though: the calibration of the positive and negative lepton momenta 
in terms of  the functions  $\epsilon_{{\rm l}^+}(\rho_{\rm l} ,  \etal  )$ 
and $\epsilon_{{\rm l}^-}(\rho_{\rm l} ,  \etal  )$ which specify the relation between 
the true and the reconstructed radius $\rho$ of track curvature 
in the magnetic field of the respective spectrometer:  
\begin{equation}
\rho _{{\rm l}^\pm}^{\rm rec}  =  \rho _{{\rm l}^\pm}^{\rm true}
\left[ 1 + \epsilon_{{\rm l}^\pm} (\rho_{\rm l} ,  \etal  ) \right]. 
\label{eq:rho}
\end{equation}
While a dependence of the calibration functions on the azimuthal angle can be factorized out
and experimentally corrected,
and therefore their possible $\phi$-dependence has been dropped, the 
dependence on $\rho_{\rm l}$ and \etal\ cannot, and must be experimentally determined from the data concurrently with the measurement of the electroweak parameters.

The above cross-sections are interpreted in terms of Standard Model parameters 
and two-dimensional PDFs.
These are for \Wpm\ 
\begin{displaymath}
u, d, s, c, \ubar ,\dbar , \sbar , \cbar , \MW , \GamW    ,
\end{displaymath}
and for Z
\begin{displaymath}
u, d, s, c, b, \ubar, \dbar, \sbar, \cbar, \bbar,  \MZ, \GamZ, \sin^2 \theta_{\rm W}, \alpha  ,
\end{displaymath}
where \MW , \MZ , \GamW\ and \GamZ\ denote the masses and widths of W and Z (assuming
$\MWp = \MWm = \MW$ and $\GamWp = \GamWm = \GamW$), $\sin^2 \theta_{\rm W}$ the electroweak mixing angle, and $\alpha$ the fine-structure constant. CKM matrix elements are considered as constants and dropped for reasons of simplicity. The  $u,d,s,c,b, \ubar, \dbar,\sbar,\cbar,\bbar$ denote two-dimensional PDFs and refer to $Q^2 = M^2_{\rm W}$. Their
evolution from $\MW^2$ to $\MZ^2$ is determined from the data as will be discussed
in Section~\ref{QSQEVOLUTION}.

Since all QCD terms, both perturbative and nonperturbative, 
will be determined experimentally when relating W and Z observables,
the accuracy of the leading and higher-order terms in the 
functional forms of the cross-sections is of secondary 
importance\footnote{One of the motivations for this approach is the non-existence of a Monte Carlo generator that provides a full representation of the interplay between QCD and electroweak effects.}.   
The functional forms of cross-sections in terms of the parameters of the Electroweak Standard Model are the ones implemented in the \WINHAC{} and \ZINHAC{}  
generators~\cite{PlaczekJadach,CarloniCalame:2004qw,Gerber:2007xk,Bardin:2008fn,
Placzek:2009jy,SiodmokPlaczek}.
\begin{table}[h]
\begin{center}
\begin{tabular}{|c|c|}
\hline
   & $\Delta \MW $ \\ 
\hline
\hline
$b^{\rm bias}$ = $1.2 \,b$ & $+42$~\MeVcsq\  \\
$b^{\rm bias}$ = $0.8 \,b$ & $-39$~\MeVcsq\ \\
\hline  
\end{tabular}
\vspace*{2mm}
\caption{Biases from uncertainties in the 3rd quark family.}
\label{WMbias3}
\end{center}
\end{table}

This approach differs from the one used at the Tevatron:  there, the measurement of the Standard Model parameters relies on one-dimensional \kT -integrated
PDFs~\cite{CTEQ6.1:2003, Martin:1994kn}, and on perturbative-QCD based algorithms for the relationship of the \pT\ distributions 
of W and Z~\cite{Aaltonen:2007ps, Abazov:2009cp}.

\subsection{Four observables}
\label{OBSERVABLES}

With a view to minimizing systematic measurement errors, the observables should have little sensitivity to detection acceptances and efficiencies. For their use in the precision measurement of several electroweak parameters, the dependence of the observables on these should be as uncorrelated as possible. Further, the observables should lend themselves to the
experimental determination of  
perturbative and nonperturbative QCD effects, 
and should clearly point to missing input if needed.

The following four ratios are proposed as observables:
\begin{eqnarray}
 \FlatAsymW (\pTl, \etal) = \frac  {\Sigma_{{\rm W}^+}(\pTl , \etal  ) - 
 \Sigma_{{\rm W}^-}(\pTl , \etal  )} {\Sigma_{{\rm W}^+}(\pTl , \etal  ) + 
 \Sigma_{{\rm W}^-}(\pTl , \etal  )}  \; ,
 \label{Wasym}
\end{eqnarray}
\begin{eqnarray}
  \FlatAsymZ (\yll, \pTll,  \pTl, \etal) = \frac  {\Sigma_{{\rm Z}^+} (\yll, \pTll, \pTl , \etal  ) - 
  \Sigma_{{\rm Z}^-} (\yll, \pTll, \pTl , \etal  )} {\Sigma_{{\rm Z}^+} (\yll, \pTll, \pTl , \etal  ) + 
  \Sigma_{{\rm Z}^-} (\yll, \pTll, \pTl , \etal  )} \; ,
  \label{Zasym}
\end{eqnarray}
\begin{eqnarray}
  \RWZ (\pTl, \etal) = \frac  {\Sigma_{{\rm W}^+} (\pTl , \etal  ) + 
  \Sigma_{{\rm W}^-} (\pTl , \etal  )} {\Sigma_{{\rm Z}^+} (\pTl , \etal  ) + 
  \Sigma_{{\rm Z}^-} (\pTl , \etal  )} \; ,  {\rm and}
  \label{RWZ}
\end{eqnarray}
\begin{eqnarray}
  \RZnorm (\pTll, \yll) = \frac {\Sigma_{\rm Z} (\pTll , \yll )} {\Sigma_{{\rm l}^+ {\rm l}^- }^{\rm norm}},
  \label{RZnorm}
\end{eqnarray}
where
\begin{eqnarray}
  \Sigma_{\rm Z} (\pTll , \yll ) = 
  \int_{M_{\rm Z} - 3 \Gamma _{\rm Z}}^ {M_{\rm Z} + 3 \Gamma _{\rm Z}} 
  \Sigma_{\rm Z} (\Mll, \pTll , \yll ) {\rm d} \Mll  \; ,                            \label{minteg}
\end{eqnarray} 
\begin{eqnarray}
 \Sigma_{{\rm Z}^{+(-)}} (\yll, \pTll, \pTl , \etal  ) & = 
 & \int_{M_{\rm Z} - 3 \Gamma_{\rm Z}}^ {M_{\rm Z} + 3 \Gamma_{\rm Z}} 
 \left[ \Sigma_{{\rm Z}^{+(-)}} (\Mll, \yll, \pTll, \pTl , \etal  ) \right] {\rm d} \Mll \; ,  {\rm and}
 \label{eq:OZ1+(-)} 
\end{eqnarray}
\begin{eqnarray}
 \Sigma_{{\rm l}^+ {\rm l}^- }^{\rm norm} = \int  \int  \int 
 \Sigma_{\rm Z} (\Mll, \pTll , \yll ) {\rm d} \Mll {\rm d} \pTll {\rm d} \yll   \; . 
 \label{norm}
\end{eqnarray} 

The latter integral is over the phase space of ${\rm l}^+{\rm l}^-$ pairs
with a back-to-back
configuration in the transverse plane, in peripheral proton--proton interactions, 
as developed and detailed in Ref.~\cite{Krasny:2006xg}.
The measurement of electroweak parameters primarily rests on lepton pairs with
their invariant mass restricted to the peak region of \MZ . 
Lepton pairs with invariant mass below the peak region of \MZ\ will play a different r\^{o}le.
They will allow to determine $Q^2$-scale effects when  
relating cross-sections at the $M_{\rm Z}^2$ scale to the ones at the $M_{\rm W}^2$ scale
(for a detailed discussion, see Section~\ref{QSQEVOLUTION}).

Although each of the proposed observables 
depends, in general, on all electroweak parameters, the respective sensitivity is different.   
The \FlatAsymW\ observable is primarily sensitive to $\MWp - \MWm$ and 
$\Gamma _{{\rm W}^+} - \Gamma _{{\rm W}^-}$.
The \FlatAsymZ\ observable is merely sensitive\footnote{With fixed values 
of \MZ\ and $\Gamma _{\rm Z}$, assumed throughout this paper.} to the value 
of $\sin^2\theta_{\rm W}$. 
The \RWZ\ observable is primarily sensitive to \MW ,  
but shows also a non-negligible sensitivity to    
$\Gamma _{\rm W}$.

All the above observables are correlated via common PDFs  
and QCD algorithms.
The  \FlatAsymW\ observable is primarily sensitive to the difference $u-d$, 
both for valence and sea quarks,
and (to a lesser extent) to the difference $s-c$.
The \FlatAsymZ\ observable is sensitive mainly to the differences between valence and sea quarks, regardless of their flavour. 
The  \RWZ\ observable is sensitive to the $u-d$ and $s-c$ differences, to the differences between the density functions of valence and sea quarks, and to the density function of b quarks.  
The \RZnorm\ observable is primarily sensitive to quark-mass effects both in the 
longitudinal and the transverse momentum density functions of the quarks.

\subsection{Lepton momentum calibration} 
\label{CALIBRATION}

Primarily to measure the \MWp\ and \MWm\ masses with a precision of 10~\MeVcsq ,  
the lepton momentum calibration functions $\epsilon _{{\rm l}^+} (\rhol,  \etal  )$ 
and $\epsilon _{{\rm l}^-} (\rhol,  \etal  )$ (see Section~\ref{CROSS-SECTIONS})  
ought to be known with the rather demanding accuracy of 
$2 \times 10^{-4}$~\cite{mW-ATLAS}. 

The proposal presented here aims first at a reduced sensitivity of electroweak parameters on
the average lepton-momentum scale, i.e. 
$\epsilon _{{\rm l}^+} (\rhol,  \etal  ) + \epsilon _{{\rm l}^-}(\rhol,  \etal  )$.  
This is advantageous, for the number of Z events will be statistically limited when
subdivided into bins of  $\rhol$ and $\etal$, and split into time intervals.
The second aim is, for the inadequacy of a charge-blind analysis at the LHC,
a calibration of the momentum-scale difference between positive and negative leptons,
i.e.  $\epsilon _{{\rm l}^+}(\rhol,  \etal  ) - \epsilon _{{\rm l}^-}(\rhol,  \etal  )$.

As for the first aim, the \RWZ\ observable is mostly concerned, and to a lesser extent the \RZnorm\ observable, while the \FlatAsymW\ and \FlatAsymZ\ observables
only weakly depend on the average lepton-momentum scale. Therefore, the discussion focuses on the \RWZ\ observable. 

The proposal of an LHC-specific calibration procedure~\cite{Krasny:2007cy} is the following.
\vspace*{1mm} 
\begin{enumerate} 

\item 
Collect data at two centre-of-mass energies  
 $\sqrt{s_1}$ and $\sqrt{s_2}= (\MZ /\MW ) \times \sqrt{s_1}$.
These two settings ascertain the same momentum fractions of the quarks 
that annihilate to W and Z, if the W sample is collected at $\sqrt{s_1}$ and the Z sample 
at $\sqrt{s_2}$. 

\item
Reduce the current $i$ of the spectrometer magnet 
when running at the lower centre-of-mass energy 
$\sqrt{s_1}$ by a  factor of $\MW /\MZ $, with a view to equalizing 
the radius of curvature \rhol\ for charged leptons  
from W and Z decays.

\item 
Use a modified version of the \RWZ\ observable defined as follows:
\begin{eqnarray}
 \RWZmod (\rhol, \etal ) = \frac  {\Sigma_{{\rm W}^+} (\rhol , \etal; s_1, i(s_1)  ) + 
 \Sigma_{{\rm W}^-} (\rhol , \etal; s_1, i(s_1)  )} {\Sigma_{Z^+} (\rhol , \etal; s_2, i(s_2)  ) + 
 \Sigma_{Z^-} (\rhol , \etal ; s_2, i(s_2)  )} \; .
  \label{RWZc}
\end{eqnarray}

\end{enumerate}
\vspace*{1mm}

The integrated luminosity  at the reduced centre-of-mass energy can be smaller than the one
at the nominal energy by a factor of ten, with a view to achieving comparable statistics of W and Z events.

It is shown in Ref.~\cite{Krasny:2007cy} and summarized in Table~\ref{tab:Syseps}, that with the use of the \RWZmod\
rather than the \RWZ\ observable, the sensitivity of the W mass measurement on the average lepton-momentum
scale is reduced by two orders of magnitude. 
This very significant gain results from the same topology of lepton 
tracks in the two settings with different
centre-of-mass energy.
\begin{table}
\begin{center}
\begin{tabular}{|c|c|c|}
\cline{2-3}
\multicolumn{1}{c|}{ } & \multicolumn{2}{c|}{$\Delta M_{\rm W}$ [\MeVcsq ] } \\
\hline
\hline
Lepton momentum bias & with \RWZ  & with \RWZmod      \\
\hline
$\epsilon_{{\rm l}^+} = +\epsilon_{{\rm l}^-}$ =$+0.005$ & $+226$ & $+5$   \\ 
$\epsilon_{{\rm l}^+} = +\epsilon_{{\rm l}^-}$ = $-0.005$ & $-223$ & $-2$   \\ 
$\epsilon_{{\rm l}^+} = -\epsilon_{{\rm l}^-}$ = $+0.005$ & $+40$ & $+22$   \\ 
$\epsilon_{{\rm l}^+} = -\epsilon_{{\rm l}^-}$ = $-0.005$ & $-19$ & $-31$   \\ 
\hline
\end{tabular}
\vspace*{3mm}
\caption{Systematic shifts of \MW\ caused by lepton momentum biases as 
defined in Eq.~\ref{eq:rho}; the statistical error
of \MW\ is 7~\MeVcsq , corresponding to an integrated luminosity of 10~fb$^{-1}$.}  
\label{tab:Syseps}
\end{center}
\end{table}

Next, the calibration of the momentum-scale difference 
$\epsilon _{{\rm l}^+}(\rhol,  \etal  ) - \epsilon _{{\rm l}^-}(\rhol,  \etal  )$ 
between positive and negative leptons is discussed. The needed accuracy is
$2 \times 10^{-3}$ if the W mass is to measured with a precision of 10~\MeVcsq\ with 
the assumption $\MWp = \MWm$, and $2 \times 10^{-4}$ if the \MWp\ and \MWm\ masses are
measured separately, as shown in Refs.~\cite{Fayette:2008wt,Fayette:2009zi}.
This demanding accuracy at the LHC contrasts with the Tevatron case, where the 
possibility of a charge-blind analysis eliminates the need of a precise calibration of the momentum-scale difference between positive and negative leptons.

The  `Double Asymmetry' method, discussed in  Refs.~\cite{Fayette:2008wt,Fayette:2009zi},
requires two running periods with opposite polarity of the spectrometer magnet. It makes use of the following modification of the \FlatAsymW\ observable:
\begin{equation}
  \DAsym_{\rm W} (\rhol) \; = \; \frac{1}{2} 
  \left[ \Asym_{\rm W}^{\vec B =  B\,\vec e_{\rm z}} \left( \rhol \right) + 
          \Asym_{\rm W}^{\vec B = -B\,\vec e_{\rm z}} \left( \rhol \right)  \right]  \; ,
\label{eq_def_dble_charge_asym}
\end{equation}
where \rhol\ represents the radius of the lepton track, and $B$ the magnetic field strength.

This latter method can be used 
for the W mass measurement and provides the needed calibration precision. 
However, as far as only the measurement of the W mass under the assumption  
$\MWp = \MWm$ is concerned,  what is needed can be obtained in a simpler 
way without changing the magnet polarity. 
Use is made of ${\rm l}^+{\rm l}^-$ pairs with invariant mass close to the Z peak.
Thanks to Nature's choice of $\sin^2 \theta_{\rm W}$ close to 1/4,   
the difference of  the \pT\ distributions of positive and negative leptons is minimal, while
the statistics of events is large. The comparison of the \Zp\ and \Zm\ \pT\ distributions
delivers what is wanted. The sensitivity to the precise value of $\sin^2\theta _{\rm W}$ is sufficiently weak to permit to factorize out the calibration procedure at the precision level of
$2 \times 10^{-3}$. 

As this precision is not sufficient for separate precision measurements of
\MWp\ and \MWm , the `Double Asymmetry' method is indispensable for these.

The conclusion is that the momentum scale of 
both positive and negative leptons can be calibrated with sufficient 
precision so as not to limit the precision of electroweak parameters.

\subsection{$\mathbf{Q^2}$ evolution} 
\label{QSQEVOLUTION}

The rationale to deal with $Q^2$ evolution builds on the concept of rescaling the LHC energy and of the field of the spectrometer magnet in such a way that production and leptonic decays
of W and Z are on the same footing: for a given W or Z rapidity, the fractions
of the proton momentum carried by annihilating quarks are the same, as is the 
radius of curvature of leptons from W and Z decays. 

The equality of the \rhol\ and \etal\ distributions of the leptons 
holds exactly, though, only for collinear massless quarks with flavour-independent
PDFs.
Even in such an ideal case the 
observables proposed in this paper are still sensitive to the $Q^2$-dependence  
of the two-dimensional PDFs
of the annihilating quarks.  
Moreover,  the \RWZmod\ observable is sensitive
to the relative normalization of the W and Z samples obtained in separate settings.

The $Q^2$-scale dependent effects concern primarily the \RWZmod\ observable, hence
the following discussion refers to this observable. The generalization to 
other observables is straightforward. 

It is proposed to select pairs of opposite-charge leptons and calculate the ratio
\begin{equation}
\ratC_{\rm QCD} = 
\frac{\int_{M_{\rm Z} - 3\Gamma _{\rm Z}}^{M_{\rm Z} + 3\Gamma _{\rm Z}} 
N_{{\rm l}^+{\rm l}^-}(s_2, i(s_2), M_{{\rm l}^+{\rm l}^-} ) \; {\rm d} M_{{\rm l}^+{\rm l}^-}} 
 {\int_{M_{\rm W} - 3\Gamma _{\rm W}}^{M_{\rm W} + 3\Gamma _{\rm W}} 
 f_{\rm BW} (s_{{\rm l}^+{\rm l}^-}, M_{\rm W} ,\Gamma _{\rm W}) \; w_{\rm EW} 
 \;  N_{{\rm l}^+{\rm l}^-}(s_1,i(s_1),M_{{\rm l}^+{\rm l}^-}) \; {\rm d} M_{{\rm l}^+{\rm l}^-}}
\label{eq:CQCD}
\end{equation}
as a function of \rhol\ and \etal\ of a randomly chosen \lp\ or \lm . 
Each pair with an invariant mass 
$M_{\rm Z} - 3\Gamma _{\rm Z}  \leq M_{{\rm l}^+ {\rm l}^-}  
\leq M_{\rm Z} + 3\Gamma _{\rm Z}$  and 
$M_{\rm W} - 3\Gamma _{\rm W} \leq M_{{\rm l}^+ {\rm l}^-}
\leq M_{\rm W} + 3\Gamma _{\rm W}$, respectively,  is weighted with the 
Breit--Wigner function\footnote{This formula corresponds to the so-called fixed-width scheme, 
however it can also be applied to the running-width scheme in which case 
both $M_{\rm W}$ and $\Gamma _{\rm W}$ have to be divided by the factor 
$\sqrt{1 + (\Gamma_{\rm W}/M_{\rm W})^2}$.} 
\begin{equation}
f_{\rm BW} (s_{{\rm l}^+{\rm l}^-}, M_{\rm W} ,\Gamma _{\rm W}) =  \frac{1}{\pi}\;
\frac{M_{\rm W}\Gamma _{\rm W}}{(s_{{\rm l}^+{\rm l}^-} - M_{\rm W}^2)^2 + 
M_{\rm W}^2 \Gamma _{\rm W}^2}\ \; ,
\end{equation}
where $s_{{\rm l}^+{\rm l}^-} = M_{{\rm l}^+{\rm l}^-}^2$.
The factor $w_{\rm EW}$ normalizes the integral of the $M_{{\rm l}^+{\rm l}^-}$ 
spectrum between  $M_{\rm W} - 3\Gamma _{\rm W}$ and $M_{\rm W} +3\Gamma _{\rm W}$  to the cross section of a $Z$-like boson with the mass and the width of the
W-boson. As a result, the numerical value of $\ratC_{\rm QCD}$ is close to unity.

In order to eliminate the $Q^2$-scale
 dependence of the two-dimensional PDFs,
the \RWZmod\ observable is replaced by the observable 
\begin{equation}
\RWZQCD (\rhol, \etal) = \RWZmod (\rhol, \etal) \times \ratC_{\rm QCD} (\rhol, \etal ) \; .
\label{eq:RWZQCD}
\end{equation}

The detector-level  simulation and the numerical evaluation of this concept is presented in 
Refs.~\cite{Krasny:2007cy}. Here, the result for the
most sensitive electroweak parameter, the W mass, is summarized in Table~\ref{tab:SyskT}.
When varying the \kT\ of quarks\footnote{Specifically: the value of 
the \Pythia{} smearing parameter $\sigma _{k_{\rm T}}$  of the flavour-independent 
partonic \kT\ density function.} over the rather conservative range 0--6~\GeVc ,
the W mass varies with the \RWZ\ observable between $-180$ and $+206$~\MeVcsq,
while there is no significant variation with the \RWZQCD\ observable. 

The \RWZQCD\ observable is  insensitive 
to the precision of the relative normalization of the two data sets taken 
at the energies $\sqrt{s_1}$ and $\sqrt{s_2}$.
\begin{table}
\begin{center}
\begin{tabular}{|c|c|c|c|}
\cline{2-4}
\multicolumn{1}{c|}{ } & \multicolumn{3}{c|}{$\Delta M_{\rm W}$ [\MeVcsq ] } \\
\hline
\hline
$\sigma_{k_{\rm T}}$ [\GeVc ]  & with \RWZ  & with \RWZmod  & with \RWZQCD     \\
\hline
0   & $-180$   & $-26$  & $-8$   \\
3   & $-68$     & $-7$    & $-3$   \\
6   & $+206$ & $+12$ & $+4$  \\
\hline
\end{tabular}
\vspace*{3mm}
\caption{Systematic shifts of \MW\ caused by different quark \kT 's; the statistical error
of \MW\ is 7~\MeVcsq , corresponding to an integrated luminosity of 10~fb$^{-1}$.}  
\label{tab:SyskT}
\end{center}
\end{table}

\subsection{The missing input}
\label{MISSINGINPUT}

At the LHC, the number of observables, Eqs.~\ref{Wasym}--\ref{RZnorm}, is four whereas {\it a priori\/} the number of
two-dimensional PDFs (for five quark flavours u, d, s, c, b) is ten = five (quark flavours) $\times$
two (quarks and antiquarks). Both the observables and the
PDFs are two-dimensional functions of one longitudinal and one transverse variable.     

The method, discussed in Section~\ref{QSQEVOLUTION}, of determining experimentally the
$Q^2$ evolution of the observables from $\MW^2$ to
$\MZ^2$, permits to define all PDFs at the  $\MW^2$ scale. For simplicity, their 
$Q^2$ dependence is henceforth dropped. 

Can the number of PDFs be reduced to match the number of observables?

In a first step it is discussed how to determine, 
at fixed $x$, the ten
$k_{\rm T} (x)$ densities for the u, d, s, c, b quarks and antiquarks. Then, in a second step, it
is discussed how to determine the $x$ dependence of the ten \kT -integrated PDFs.

It can be assumed that, at fixed $x$, the \kT\ dependence of the two-dimensional
PDFs of quarks and antiquarks is the same. This is suggested because W and Z bosons
are produced predominantly by the annihilations of sea quarks, hence 
the equality  $k_{\rm T}^{\rm q} (x) = k_{\rm T}^{\bar{\rm q}} (x)$ is a reasonable assumption
for all five quark flavours\footnote{The contributions from valence quarks calls for a small
correction to be applied.}. This reduces the number of needed
\kT\ densities from ten to five.
Assuming further that the \kT\ densities are the same for the u and d quarks, 
the \pT\ dependences of the four LHC observables permit the determination of all four
remaining \kT\ densities, $k_{\rm T}^{{\rm u},{\rm d}} (x)$, $k_{\rm T}^{\rm s} (x)$,
$k_{\rm T}^{\rm c}(x)$, and $k_{\rm T}^{\rm b}(x)$. In practice, better precision is obtained
if in addition the further assumption $k_{\rm T}^{{\rm u},{\rm d}} (x)$ = $k_{\rm T}^{\rm s}(x)$
is made. 

As for the $x$ dependence of the ten needed \kT -integrated PDFs, 
the equality of the quark and antiquark densities of the s, c, and b flavours 
is assumed\footnote{For s quarks, a small violation of this equality
is likely which calls for a small correction to be applied;  however, for u and d quarks,
at $x \sim 6 \times 10^{-3}$ such equality is violated at the level of 
$\sim$15\%.}, $s(x)=\sbar(x)$,  $c(x)=\cbar(x)$ and 
$b(x)=\bbar(x)$. This reduces
the number of needed \kT -integrated PDFs from ten to seven.
Given the $\eta$ dependences of the four LHC observables, 
three experimental constraints are missing, sufficiently precise input PDFs are needed from elsewhere.

The following three  \kT -integrated flavour-singlet PDFs are least constrained  
by the LHC data alone: $\uv(x)  - \dv(x)$, $s(x) - c(x)$ and $b(x)$. 
They are referred to below as `missing input'. 
This choice of missing input is also made in other papers 
on this subject (Refs.~\cite{Krasny:2007cy, Fayette:2008wt, Upcoming_GammaW, Upcoming_SIN}). 

The rationale behind the choice of the bulk of the missing input in terms of differences
of PDFs is to focus attention on the possibility of compensating PDF changes that was 
discussed in Section~\ref{BIASEDWMASS}.

The reason for the preference of the flavour-nonsinglet PDF $\uv  - \dv$
is that it will have to be obtained from data taken at smaller $Q^2$ scales 
and subsequently extrapolated to the $\MW^2$ scale.
The $Q^2$ evolution of non-singlet PDFs is independent of the initial form of the gluon 
density function, hence the extrapolation uncertainty is reduced. 

The crucial point is whether the missing input, taken from existing data, is precise enough.

The uncertainties in the missing input used in the pertinent 
studies in Refs.~\cite{Krasny:2007cy, 
Fayette:2008wt, Upcoming_GammaW, Upcoming_SIN} reflect the 
uncertainties of the experimental data used to obtain the missing input. 
The studies gave the following results: an 
uncertainty of $\cal{O}$(100)~\MeVcsq\ 
for \MW\ and for the difference $\MWp - \MWm$, an uncertainty of $\cal{O}$(40)~\MeVcsq\ 
for  $\Gamma_{\rm W}$ , and an uncertainty of
$\cal{O}$(0.001) for $\sin^2\theta _{\rm W}$. 
Already for an integrated luminosity as small as 1~fb$^{-1}$ the errors that result from the uncertainties of today's missing input, are larger than statistical and systematic 
errors stemming from the LHC data.   

The conclusion is that the current precision~\cite{Amsler:2008zzb} of the above electroweak parameters cannot be improved at the LHC unless the precision of the missing input is
significantly improved. This conclusion is in conflict with the prognoses made by the 
LHC experiments. The conflict is particularly apparent for the W mass where the measurement precision is found to be 5--10 times worse than estimates made by the LHC 
Collaborations~\cite{mW-ATLAS, mW-CMS}.   

In order to ascertain the origin of these discrepancies the analysis was repeated 
using the LHC Collaborations' method of studying the impact of the uncertainties of 
\kT -integrated PDFs on the electroweak observables. With the same range
of uncertainty of the parameters of the \kT -integrated PDFs, the results became compatible
with the results of the LHC Collaborations. 

The discrepancy can be traced back to two sources: a lack of considering
compensating PDF changes especially in regions where such changes are hardly
constrained by existing experimental data, and  
too rigid a restriction of the functional 
forms of the missing input at the initial hardness scale.

\section{Ways forward}
\label{WAYSFORWARD}

There is much discussion about improvements of the parton density functions 
from HERA experiments.
The HERA programme is completed. Results from the e$^\pm$--proton scattering
data of the H1 and ZEUS Collaborations have been published~\cite{H1:2009wt},
more results from common analyses are forthcoming. Although the
ultimate measurement errors are expected to be reduced by a factor of up to two,
the level of uncertainty of the PDFs of u$_{\rm v}$, d$_{\rm v}$ and s 
quarks as assumed in this paper's analysis is appropriate. 
The HERA data are dominated by neutral-current e$^\pm$--proton scattering, while the 
separation between quarks and antiquarks requires charged-current scattering.  The scarce
statistics of charged-current scatterings (less than 20 k events) render them inadequate to provide the missing input for the LHC. Moreover, the neutral-current scatterings are largely insensitive to compensating changes of the PDFs of \uv\ and \dv\ quarks. 

The final results of the H1 and ZEUS Collaborations on heavy-flavour production are not yet available. However, again for scarce statistics, these data cannot pin down 
the PDFs of c and b  quarks relative to those of the u and d quarks at the required level of
$\sim$1\%.

Also, the present and the possible future experimental programme 
at the Jefferson Laboratory cannot improve the knowledge 
of the proton PDFs at the $\MW^2$ and $\MZ^2$ scales. 
This is because only a fraction of the pertinent deep-inelastic scattering 
data---where the higher-twists and target-mass 
corrections can be neglected---lends itself to 
QCD fits of PDFs\footnote{For example, in the MSTW set of
QCD fits, only data are used that satisfy the condition  
$W^2 > 15$~GeV$^2/c^4$ on the squared hadronic mass~\cite{MSTW}.}. 
At the Jefferson Laboratory  where $W^2_{\rm max} = 11$~GeV$^2/c^4$, 
the relevant kinematical region is beyond reach.

If, as planned, 
the electron beam momentum at the Jefferson Laboratory is increased to 12~GeV/{\it c}, 
the boundary of the useful region will be crossed but only
barely so\footnote{The useful data would have inelasticity $y > 0.75$ 
where resonant photo-production processes are dominant 
and where QED radiative corrections are large.}.

\subsection{Deuteron--deuteron collisions at the LHC}
\label{DEUTERONS}

The impact of the uncertainties from missing input PDFs
can be considerably reduced by operating the LHC with isoscalar beams.
The natural choice is to collide deuteron beams.  

The LHC luminosity is expected to scale like $\cal{L}_{\rm ion-ion} = \cal{L}_{\rm pp}\,$/$A^2$
where $A$ is the mass number of a light ion. Then the event rates with high 
\pT -signatures will be comparable for the proton and for the light-ion collisions. 
The experimental environment at the LHC that is characterized by  
multiple proton--proton collisions within the same bunch-crossing,  
is not rendered more difficult by parasitic collisions of spectator nucleons 
in light-ion collisions. 

The deuteron beams restore isospin symmetry for the quarks of the 1st family.
The four independent  \kT -integrated PDFs
$u(x)$, $d(x)$, $\ubar(x)$ and $\dbar(x)$ are reduced to two: 
$u(x)+d(x)$ and $\ubar(x) +\dbar(x)$.
Equality of \Wp\ and \Wm\ production is restored and the spin-density matrices of W and Z produced by quarks of the 1st family are nearly the same\footnote{It is assumed that the 
$Q^2$-evolution from the $\MW^2$ to the $\MZ^2$ scale is handled as proposed in
Section~\ref{QSQEVOLUTION}.}.
If the contributions from quarks of the 2nd and 3rd family could be neglected, the isospin 
symmetry of deuterons at the LHC would play the same r\^{o}le as the matter--antimatter 
symmetry at the Tevatron.  

The isospin symmetry of the 1st quark family reduces the number of needed two-dimensional PDFs from ten to eight.
With the assumption that the PDFs of the s, c and the b quark flavours are the same for quarks and antiquarks, the number of needed PDFs is further reduced from eight to five. 
Given the four
constraints from the measured \Wp\ and \Wm\ and Z 
cross-sections in \dd\ collisions,
there is only one two-dimensional PDF left unconstrained.  

With a view to solving this problem, the sensitivity of the most sensitive electroweak parameter,
the W mass,  
to the uncertainty in the b quark density function has been investigated by analyzing the Z
cross-sections not in the full pseudorapidity range $| \etal | < 2.5$, but 
in the restricted region $2 < | \etal | < 2.5$. Since the contribution of b\={b} annihilations 
is reduced in this kinematical region, the sensitivity of the Z cross-section to the b quark PDFs is reduced, too.
Varying the b-quark PDF  by 40\% from its central  value,  the W mass changed by 
5~\MeVcsq , comparable with the statistical error of the pseudo-data sample. 
Since this is perfectly
acceptable, one might conclude that taking data with deuterons in the LHC would provide the
wanted precision of electroweak parameters.

However, caveats remain.

A limitation arises from the statistical error of 
the $\FlatAsymW$ observable that measures in \dd\ collisions directly 
the $s(x)-c(x)$ distribution~\cite{Fayette:2008wt,Fayette:2009zi}.
For the smallness of the Cabibbo angle,  reducing the statistical error to the 
level sufficient to determine \MW\ with a precision of 10~\MeVcsq\ 
requires a substantial integrated luminosity of \dd\ collisions: 25~fb$^{-1}$. 
 
The PDFs  of the proton and of the neutron bound in deuterons are different with respect to the PFDs of free nucleons. The nuclear binding effects, off-shellness, and shadowing 
effects, could however be absorbed 
in a consistent way into the W and Z observables proposed in Section~\ref{OBSERVABLES}.

In summary, high-statistics data from \dd\ collisions at the LHC would be
sufficient to provide electroweak parameters with the desired precision.

\subsection{\pp\ at the LHC, \ppbar\ at the Tevatron, and muon--nucleon scattering combined}
\label{MUONSCATTERING}

The concept of solving the missing-input problem by \dd\ collisions in the LHC is elegant and
technically feasible, though not realistic in the near future. Therefore, an alternative
is discussed: obtaining with sufficient precision 
from a joint analysis of Tevatron \ppbar\ data, of data 
from a new muon--nucleon scattering experiment, and of LHC \pp\ data,
all needed PDFs with adequate precision.

Two intrinsic difficulties come along with this concept.

A minor difficulty is cross-normalization between \pp\ and \ppbar\ experiments with
adequate precision\footnote{The majority of the observables are 
normalization-independent ratios
but not all, see Eq.~\ref{RZnorm}.}. This problem can be solved by measuring the luminosity
through the well-known cross-section of ${\rm l}^+{\rm l}^-$ pairs with a back-to-back
configuration in the transverse plane,
in peripheral proton--proton interactions and proton--antiproton interactions,
respectively~\cite{Krasny:2006xg}.

A major difficulty is that very different $x$ domains of
pertinent PDFs are populated, and that for W and Z production only the product
of the $x$ values of the annihilating quarks is known, $x_1 \cdot x_2$. Therefore,
different $x$ intervals must be considered, ranging from
`very small $x$' at the LHC over `small $x$' at the Tevatron until `medium $x$'
at the muon--nucleon scattering experiment. The intervals
are partially overlapping---generally, if $x_1$ is small, $x_2$ is large, and {\em vice versa\/}.
In the following it is discussed how in a coherent way,
yet with different strategies in different $x$ intervals, the information from different
intervals can be linked and the missing high-precision PDFs for the analysis 
of LHC \pp\ data across the full range of $x_1$ and $x_2$ be obtained.

The LHC is operated at a much higher energy than the Tevatron. 
Be $x^{\rm LHC} _{\rm low}$ and $x^{\rm LHC} _{\rm high}$
the minimal and the maximal $x$ of the quarks that produce W and Z at the LHC, and
$x^{\rm TEV} _{\rm low}$ and $x^{\rm TEV} _{\rm high}$ their equivalents at the Tevatron.
The unfolding of the PDFs will be different in the LHC-exclusive interval [$x^{\rm LHC} _{\rm low} , 
x^{\rm TEV} _{\rm low}$], in 
the overlap interval [$x^{\rm TEV} _{\rm low} , x^{\rm LHC} _{\rm high}$], and in 
the Tevatron-exclusive interval [$x^{\rm LHC} _{\rm high} , x^{\rm TEV} _{\rm high}$].

It is assumed that the observables have been 
corrected for their $Q^2$ evolution and are defined at 
$Q^2 = \MW^2$.
Likewise, the needed two-dimensional PDFs refer to $Q^2 = \MW^2$ and have
no energy dependence\footnote{A possible small energy dependence of the $k_{\rm T} (x)$
distributions, at fixed $x$,  can be corrected for.}.

Like in Section~\ref{MISSINGINPUT}, in a first step it is discussed how to determine, 
at fixed $x$, the ten
$k_{\rm T} (x)$ densities for the u, d, s, c, b quarks and antiquarks. Then, in a second step, it
is discussed how to determine the $x$ dependence of the ten \kT -integrated PDFs.

The discussion proceeds in the order of increasing $x$.
\vspace*{1mm}
\begin{itemize}

\item  In  the {\em LHC-exclusive interval\/}  which concerns the very-low $x$ region, 
the analysis problem was partly discussed already
in Section~\ref{MISSINGINPUT}. As for the $k_{\rm T} (x)$ densities,
it was concluded that  the three densities $k_{\rm T}^{{\rm u},{\rm d},{\rm s}} (x)$,
$k_{\rm T}^{\rm c}(x)$, and $k_{\rm T}^{\rm b}(x)$ can be determined from the
\pT\ dependence of the four LHC observables. 

As for the $x$ dependence of the ten needed \kT -integrated PDFs, 
the equality of the quark and antiquark densities of all five quark flavours 
is assumed, for W and Z are dominantly produced in the annihilation
of sea quarks\footnote{At $x \sim 6 \times 10^{-3}$, this assumption is violated for
the u and d quarks at the $\sim$15\% level which calls for an appropriate
correction to be applied.}. This reduces the number of needed \kT -integrated
PDFs from 10 to five, still one more than can be determined from the 
$\eta$ distribution of the four LHC observables.

The obvious choice is to rely on the `HERA-combination'  of PDFs,
\begin{displaymath}
4/9\left[u(x) + \ubar(x) + c(x) + \cbar(x)\right] + 1/9\left[d(x) + \dbar(x) + s(x) + \sbar(x)\right] \; , 
\end{displaymath}
evolved with the DGLAP equations from the HERA-$Q^2$  to the $\MW^2$ 
scale. However,
this evolution algorithm that {\it per se} leads already to a $\sim$2\% uncertainty at the
$\MW^2$ scale~\cite{H1:2009wt}, is not undisputed for 
very low $x$~\cite{GolecBiernat:1994cd}.

With the latter proviso, the problem is solved.  
  
\item In the {\em overlap interval\/}, the needed ten $k_{\rm T} (x)$ densities
are constrained by the \pT\ dependence of eight observables, four from the LHC and 
four from the Tevatron, which are sufficiently independent of each other.
Two assumptions must be made. A straightforward choice is
$k_{\rm T}^{\rm c} (x)$ = $k_{\rm T}^{\bar{\rm c}} (x)$ and 
$k_{\rm T}^{\rm b} (x)$ = $k_{\rm T}^{\bar{\rm b}} (x)$.

The same assumptions can also be made for the $x$ dependence of the 
\kT -integrated PDFs of the c and b quarks and antiquarks\footnote{In case that the 
Tevatron precision
of the $\FlatAsymZ$ observable is not good enough, the equality of the two-dimensional
PDFs of the s quarks and antiquarks can be assumed with negligible effects 
on the uncertainties of the measured electroweak parameters.}, $c(x) = \bar{c}(x)$ and
$b(x) = \bar{b}(x)$. Then from the $\eta$ dependence of the eight observables the
$x$ dependences of the eight remaining \kT -integrated PDFs can be determined. 

The problem is solved.

\item In the {\em Tevatron-exclusive interval\/},  the contribution of annihilations involving the bottom quark can be neglected, which reduces the needed \kT\ densities from ten to eight.
Assuming further that $k_{\rm T}^{\rm c} (x) = k_{\rm T}^{\bar{\rm c}} (x)$, that  
the \kT\ densities of the u, d, and s antiquarks is the same, 
$k_{\rm T}^{\bar{\rm u},\bar{\rm d},\bar{\rm s}} (x)$, and that the \kT\ densities of the
u and d valence quarks are the same, $k_{\rm T}^{\rm {u_{\rm v}},\rm {d_{\rm v}}} (x)$, 
all four remaining  \kT\ densities can be
determined from the \pT\ dependence of the four observables.

As for the $x$ dependence of the eight needed \kT -integrated PDFs that remain after
neglecting the contributions from b quarks and antiquarks, it is assumed that
$c(x) = \bar{c} (x)$ and  
$c(x) = 0\;$\footnote{Close to the overlap  interval, at  $x \sim 8 \times 10^{-2}$, the latter 
assumption is violated at the  $\sim$5\% level which requires a small correction to be
applied.}. This leaves six \kT -integrated PDFs to be determined. Given only the $\eta$
dependences of four Tevatron observables, two \kT -integrated PDFs remain unconstrained. 

There remains a problem which can be solved only with additional data. Such data 
would be provided by a high-precision muon--nucleon scattering experiment.

\end{itemize}

The muon--nucleon scattering experiment would measure from the 
deep-inelastic scattering of $\cal{O}$(100)~\GeVc\ muons on stationary hydrogen and
deuterium targets the asymmetry
\begin{eqnarray}
\cal{A}^{{\rm p},{\rm n}}_{\rm DIS} & = &\frac{\sigma (\mu,{\rm p}) - \sigma (\mu,{\rm n})}
                                                                       {\sigma (\mu,{\rm p}) + \sigma (\mu,{\rm n})} 
                                                                       \nonumber \\
                   & \propto & u_{\rm v} - d_{\rm v} + 2 \cdot (\bar{u} - \bar{d}) + {\rm corr.} \; ,
\end{eqnarray}                                                                                                                                     
which is---as far as the difference between $\bar{u}$ and $\bar{d}$
is concerned---complementary 
to what is measured by the W production asymmetry (Eq.~\ref{Wasym})
\begin{eqnarray}
\FlatAsymW & \propto & u_{\rm v} \cdot \bar{d} - d_{\rm v} \cdot \bar{u} + {\rm corr.} 
\end{eqnarray}
It is recalled that the difference between $\bar{u}$ and $\bar{d}$ is an important 
ingredient for the understanding of \Wp , \Wm and Z polarizations at the LHC.

The asymmetry $\cal{A}^{{\rm p},{\rm n}}_{\rm DIS}$ has the advantage of 
bypassing normalization problems but provides only one constraint where two
are needed. For the second constraint, 
the assumption $s(x) = \bar{s}(x)$ is a reasonable choice\footnote{This assumption
can be questioned; in this case, rather than measuring the 
asymmetry $\cal{A}^{{\rm p},{\rm n}}_{\rm DIS}$, absolute cross-sections 
$\sigma (\mu,{\rm p})$ and $\sigma (\mu,{\rm n})$ would have to measured.}.

With the inclusion of the muon--nucleon scattering data, the problem of
missing high-precision PDFs for the analysis of LHC \pp\ data is solved.

The present uncertainty on $\cal{A}^{{\rm p},{\rm n}}_{\rm DIS}$ 
from lepton--nucleon scattering experiments stems from three sources: 
(i) the statistical precision of (1--4)\%, (ii) 
uncertainties of $\sim$2\% in nuclear corrections, and (iii) uncertainties 
of $\sim$2\% in the $Q^2$ evolution to the  $\MW^2$ scale. The new muon--nucleon
scattering experiment
would have to reduce the statistical error by a factor of four, and improve 
by a comparable factor the experimental and theoretical control of uncertainties from
nuclear effects in the deuteron. As for the latter, the electron--nucleon scattering 
programme at the Jefferson Laboratory is expected to provide new insights. 
 
A Letter of Intent~\cite{LoI} for such an experiment was submitted to CERN Programme Committees. Therein, the exposure of the COMPASS detector to the 
muon beam of the CERN--SPS was proposed.

\section{Conclusion}
\label{CONCLUSION}

The measurement of the W mass at the LHC with a precision of $\cal{O}$(10)~\MeVcsq\
is {\em per se\/} important, and even more important if the Higgs boson will not be found.
However, the prognoses by the LHC Collaborations that they can achieve this precision are
much too optimistic, for the inadequate knowledge of certain 
proton PDFs that are not relevant in the analysis of \ppbar\ collisions at the Tevatron but
are relevant in the analysis of \pp\ collisions at the LHC. 

The missing input for the precise measurement of parameters of the Electroweak
Standard Model from \pp\ collisions at the LHC 
is identified. Proposals are discussed how to solve the missing-input problem.
One possibility is to complement the \pp\ programme of the LHC with a deuteron-deuteron
collision programme. Another possibility is to obtain missing input from 
a new high-precision muon--nucleon scattering experiment, and to analyze
these data coherently with LHC \pp\ and Tevatron \ppbar\ data.
In the framework of an LHC-specific programme
for the precision measurement of parameters of
the Electroweak Standard Model, a precision of 10~\MeVcsq\ of \MW\ 
can be achieved.

Unless efforts as discussed in this paper are undertaken, the  
precision of the W mass, and of other parameters of
the Electroweak Standard Model, will not be improved at the LHC. Thus a  chance may be 
missed towards understanding the mechanism that regularizes the unitarity problem of 
this Model.



%
%

\end{document}